\documentclass[fleqn,usenatbib]{mnras}
\usepackage{graphicx} 
\usepackage[utf8]{inputenc}
\usepackage{amsmath,amsfonts}
\usepackage{url}
\usepackage{hyperref}
\usepackage{cuted}
\usepackage{subcaption}
\usepackage{newtxtext,newtxmath}
\usepackage[T1]{fontenc}
\usepackage{listings}
\usepackage{xcolor}
\usepackage[ruled,vlined]{algorithm2e}
\SetKw{KwBy}{by}
 
\DeclareRobustCommand{\VAN}[3]{#2}
\let\VANthebibliography\thebibliography
\def\thebibliography{\DeclareRobustCommand{\VAN}[3]{##3}\VANthebibliography}

\title[FIP-TOI: Fast Imaging Pipeline for Pulsar Localisation]{FIP-TOI: Fast Imaging Pipeline for Pulsar Localisation with a Transient-Oriented Radio Astronomical Imager}

\author[X. Li et al.]{
Xiaotong Li,$^{1}$
Karel Ad\'{a}mek,$^{1,2}$
Olexa Bilaniuk,$^{3}$
Vladislav Stolyarov,$^{4,5}$
and Wesley Armour$^{1}$\thanks{E-mail: wes.armour@oerc.ox.ac.uk}
\\
$^{1}$Department of Engineering Science, University of Oxford, 7 Keble Rd, Oxford OX1 3QG, UK\\
$^{2}$Institute of Physics, Silesian University in Opava, Bezru\v{c}ovo n\'{a}m. 13, Opava 74601, Czech Republic\\
$^{3}$Innovation, Development and Technologies, Mila - Québec Artificial Intelligence Institute, 6666 Rue Saint-Urbain, Montréal, QC H2S 3H1, Canada\\
$^{4}$Astrophysics Group, Cavendish Laboratory, University of Cambridge, JJ Thomson Ave, Cambridge CB3 0HE, UK\\
$^{5}$Special Astrophysical Observatory of 
the Russian Academy of Sciences,
Nizhny Arkhyz 369167, Russia
}
\date{Accepted XXX. Received YYY; in original form ZZZ}
\pubyear{2025}

\begin{document}
\label{firstpage}
\pagerange{\pageref{firstpage}--\pageref{lastpage}}
\maketitle

\begin{abstract}
Rapid localisation of radio transients requires imaging pipelines that combine low latency, high sensitivity, and accurate transient positioning. Fast Imaging Pipeline (FIP) defines a (quasi)-real-time image-based transient detection pipeline for next-generation radio telescopes such as the Square Kilometre Array (SKA). However, the current FIP within the SKA Science Data Processor (SDP) remains at an early stage, leaving substantial scope to improve its performance. To address this need, we develop FIP-TOI, which combines a novel Transient-Oriented Imager (TOI) with the transient detector FITrig. FIP-TOI is highly parallelised and GPU accelerated. In experiments, TOI improves signal-to-noise ratio (SNR) by at least 15 compared with WSClean and SKA-SDP imagers, while providing higher positional precision than WSClean snapshot imaging. On a 396 GB MeerKAT Measurement Set containing 1500 snapshots, FIP-TOI processes the data in 170-190 s on a single NVIDIA H100, corresponding to approximately 120 ms per snapshot, while using 714 MB of RAM and about two CPU cores. Testing on diverse datasets --- including fields with multiple faint transients, an on-and-off pulsar, and a pulsar exhibiting intensity changes --- FIP-TOI demonstrates robust performance across all scenarios and achieves lower localisation errors than SKA-SDP-FIP.
\end{abstract}


\section{Introduction}

Celestial transients, such as pulsars \citet{Pulsar}, supernovae \citet{supnov}, and fast radio bursts (FRBs) \citet{transient2}, are unique celestial sources characterised by significant changes over short timescales. Detecting and localising these transients often rely on snapshot surveys such as ThunderKAT \citet{thunderkat}, ASKAP VAST survey \citet{askap_vast}, \texttt{realfast} VLA survey \citet{realfast}, and Fast Imaging Pipeline (FIP) in Square Kilometre Array (SKA) Science Data Processor (SDP), SKA-SDP-FIP \citet{fipska}. 

FIP, an image-based transient detection pipeline, comprises ``imaging'' and ``localisation'' components. It reconstructs images from visibilities (correlations of radio signals received by antennas) captured over short durations. These reconstructed ``snapshot'' images are then statistically analysed to identify transient candidates. In the context of a wide-field blind search, a transient signal is not necessarily the brightest feature in a snapshot image, particularly when it is embedded in fields containing bright stationary sources. As the transient location is not known \textit{a priori}, the analysis must search across the full field of view (FOV) rather than a predefined position, which motivates the use of source finders for the localisation. While both the VLA/realfast system and SKA/FIP use snapshot imaging and statistical analysis to detect transients, FIP is fundamentally designed for the scale and complexity of next-generation telescopes, such as the SKA. Moreover, FIP explicitly separates imaging and localisation, allowing each to be independently optimised for higher precision and flexibility. Notably, the term ``fast'' in FIP refers to the high-efficiency, low-latency computational design of the pipeline, rather than the timescale of the celestial transient itself.

In current FIP designs, source finding techniques are commonly used in the ``localisation'' component, such as SFIND \citet{sfind}, AEGEAN \citet{sourcefinder1,aegean1}, SOurce FInding Application (SoFiA-2) \citet{sofia22,sofia2}, and Fast Imaging Trigger (FITrig) \citet{FItrigger}. Among these, FITrig provides a real-time, low-false-positive approach capable of detecting faint transients even in the presence of bright stationary sources in the same FOV, making it particularly effective for robust transient localisation. It first produces a tLISI (transient-oriented Low-Information Similarity Index) matrix, highlighting regions likely to contain transients. These candidate regions are then passed to image- or image-frequency-domain algorithms to precisely localise the transients' positions.

Currently, astronomers employ radio astronomical imagers, such as W-Stacking Clean (WSClean; \citealt{w6}) and imaging tools in Common Astronomy Software Applications (CASA; \citealt{CASA,CASAnew}) for the ``imaging'' component \citet{realpul1,realpul2}. These imagers can reconstruct individual snapshots of the sky; however, they lack the computational efficiency needed for transient detection, particularly in real-time analysis for large radio surveys. The limitation becomes more pronounced with the vast data volumes that will be collected by next-generation telescopes, such as the SKA\footnote{\url{https://www.skao.int/en}} \citet{skaconf} and the Deep Synoptic Array (DSA) 2000\footnote{\url{https://www.deepsynoptic.org/overview}}.

Specifically, in radio interferometric imaging, image reconstruction is typically formulated as an inverse problem, where the relationship between the observed visibility and the received sky brightness distribution (SBD; \citealt{RA7}) follows the van Cittert-Zernike theorem \citet{vcz}, which can be expressed as
\begin{equation}
\mathit{Vis}\left( {u,v,w} \right) = {\iint{I\left( {l,m} \right)e^{- i2\pi{({ul + vm + w{({\sqrt{1 - l^{2} - m^{2}} - 1})}})}}\mathrm{d}l\mathrm{d}m}} ,
\label{vis}
\end{equation}
where $\mathit{Vis}$ is the observed visibility, $I$ is the received SBD, $u,v,w$ represent coordinates in the frequency domain, while $l,m,n=\sqrt{1 - l^{2} - m^{2}}$ represent coordinates in the spatial domain. The $w$ indicates the delay tracking direction, with $u$ and $v$ lying in the perpendicular plane. A dirty image is obtained by applying the adjoint of the measurement operator to the observed visibilities, yielding the sky brightness convolved with the dirty beam due to incomplete sampling of the visibility space.

A baseline is defined by the vector connecting the phase centres of two antennas in the radio interferometer. For non-coplanar baselines, the adjoint operator in wide-field imaging depends on the \textit{w}-term in Equation (\ref{vis}). Consequently, practical algorithms for computing wide-field dirty images must specify how the \textit{w}-term is incorporated into the adjoint, leading to different numerical implementations such as \textit{w}-projection \citet{w3}, \textit{w}-snapshots \citet{w4,wssnap}, and \textit{w}-stacking \citet{w6,wgridder} approaches. Each of these approaches faces challenges for real-time transient imaging. The computational and memory demands of handling the \textit{w}-term become prohibitive for the large datasets produced by next-generation transient surveys, regardless of whether convolution kernels, plane fitting, or layer-based Fast Fourier Transforms (FFTs) are used.

In this article, we develop a new imager, Transient-Oriented Imager (TOI), designed for next-generation radio telescopes, where large-scale data processing is a fundamental challenge. TOI improves computational efficiency and increases signal-to-noise ratio (SNR) in transient detection. It takes the observed visibilities in each sample time as input and generates dirty snapshots. Graphics Processing Units (GPUs) are leveraged to maximise computational throughput. 

TOI is integrated with the GPU-accelerated FITrig \citet{FItrigger} to construct FIP-TOI, an image-based transient detection pipeline designed for high-sensitivity detection of faint transients and computationally efficient processing. Here, FIP-TOI denotes the pipeline developed in this work (TOI+FITrig), and is distinct from the SKA-SDP-FIP \citet{fipska} (see Section \ref{FIPsec52} for details). Because FITrig operates directly on dirty images, the need for deconvolution is eliminated in the pipeline. The resulting pipeline, FIP-TOI, is capable of handling large-scale data in real time, achieving high throughput through distributed and parallel computing.

This article begins with an overview of the FIP design, as outlined in Section \ref{FIPsec52}. Section \ref{FIPsec53} provides a detailed description of the development of the TOI, with a focus on its underlying mathematical principles. Section \ref{FIP54} presents the parallelisation strategy for GPU acceleration of FIP-TOI. The performances of TOI and FIP-TOI are demonstrated in Section \ref{FIP55}. In Section \ref{FIP553}, we present and discuss the performance of FIP-TOI under different scenarios, using simulated and real datasets. We present our conclusions in Section \ref{FIP57}. FIP-TOI\footnote{\url{https://github.com/egbdfX/FastImagingPipe}} is open source and is available from GitHub pages.

\section{Fast imaging pipeline}
\label{FIPsec52}

In general, FIP is designed to detect and localise transient candidates with low latency, ideally in real time, enabling rapid follow-up observations and detailed downstream analysis. It operates on dirty snapshots (``imaging'') to rapidly identify statistically significant short-duration emission consistent with transient behaviour (``localisation''), which is not possible by traditional time-domain detection methods. FIP therefore targets the discovery stage of large-scale transient surveys, where the primary objective is to determine when and where a transient-like event occurs. Detailed time-domain characterisation, light-curve reconstruction, and physical classification are beyond the scope of this pipeline and are expected to be handled by downstream processing.

\begin{figure}
    \includegraphics[width=\columnwidth]{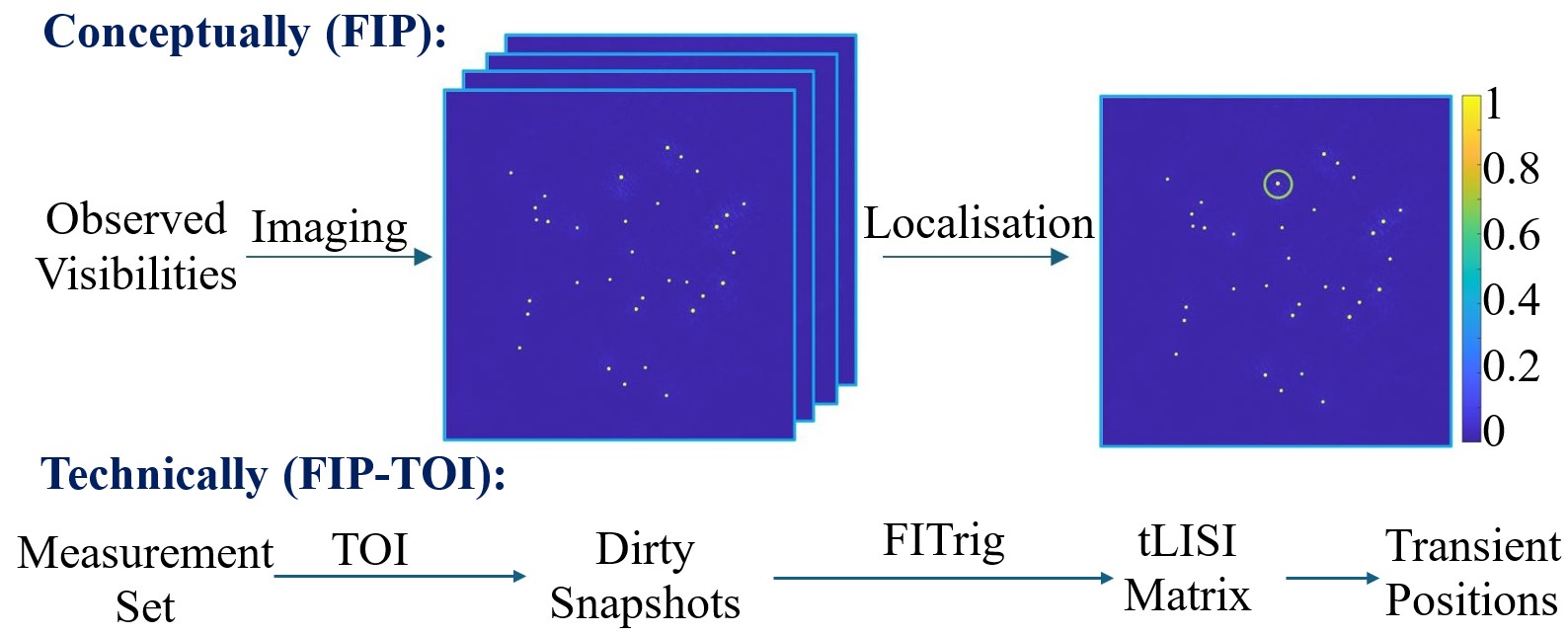}
\caption{FIP diagrams. The conceptual diagram is a general workflow shared among different FIPs, including the FIP proposed in this article (FIP-TOI) and other FIPs, such as the SKA-SDP-FIP \citet{fipska}. The technical diagram illustrates the structure of FIP-TOI. The example images are simulated with sky model based on the GLEAM dataset \citet{gleam}. In the simulation, a source is removed from some snapshots to represent a transient, as highlighted in the image on the right by the circle. Pixel intensities are represented by the colour bar. Images are normalised in this example for illustration purposes. \label{pipe}}
\end{figure}

The diagrams of FIP are shown in Fig. \ref{pipe}. The conceptual diagram is a general workflow shared among different FIPs, including the FIP proposed in this article (FIP-TOI) and other FIPs, such as the SKA-SDP-FIP \citet{fipska}. The technical diagram, on the other hand, illustrates the structure of FIP-TOI.

SKA-SDP-FIP ~\citet{fipska} is a pipeline from Processing Function Library\footnote{ \url{https://developer.skao.int/projects/ska-sdp-func/en/latest/}} (PFL) developed for image-based transient detection in SKA-scale radio surveys. PFL was developed as part of the SKA-SDP prototyping efforts to demonstrate a high-performance library capable of processing SKA-like data volumes at the quality required by SKA. PFL focused on simple, portable interfaces that enable users to utilise both CPU- and GPU-accelerated parallelised implementations of algorithms for image-domain radio astronomy from both C/C++ and Python.

SKA-SDP-FIP is organised into three main stages. The first stage constructs snapshot images from a set of visibilities, based on \textit{w}-stacking algorithm. The output of this stage consists of difference maps, generated from successive dirty images, which are expected to contain transients. The second stage performs point source detection on each difference map by SoFiA-2, considering only positive flux densities to avoid spurious detections. This stage produces a set of detection catalogues (one per difference map) which are subsequently merged into a single catalogue for further analysis. The third and final stage applies cluster analysis to the merged catalogue, identifying groups of neighbouring sources across multiple snapshots that are likely associated with the same transient event. This step helps to suppress false detections that are randomly scattered across the field. The pipeline culminates in the production of a final catalogue of transient candidates.

To improve computational efficiency and enable real-time transient detection, we develop FIP-TOI in this article. Unlike the existing SKA-SDP-FIP, which relies on sequential source detection, catalogue generation, catalogue merging, and cluster analysis, FIP-TOI identifies transient candidates directly from the temporal behaviour of image sequences. This design eliminates several intermediate processing steps, reducing computational overhead, while making the pipeline naturally suited to streaming observations. In addition, FIP-TOI is highly automatic. The mandatory inputs of FIP-TOI required from users are the Measurement Set, image size, and the size of each pixel (cell size); in comparison, in SKA-SDP-FIP, users need to configure the parameters used in SoFiA-2, according to dirty beam and noise level.

Its imaging component is TOI, a highly parallelised, high-SNR single-snapshot imager designed for transient detection. More details are presented in Section \ref{FIPsec53}. TOI provides higher computational efficiency than the \textit{w}-stacking algorithm while achieving higher imaging precision than the \textit{w}-snapshot algorithm (see Section \ref{FIP55}). It generates a dirty snapshot for each time sample from the associated visibilities in the Measurement Set.

The dirty snapshots are passed to FITrig, a statistics-based technique designed to detect and localise pulsars in radio dirty images \citet{FItrigger}. Rather than performing conventional source finding on individual difference images, FITrig identifies transient signals by analysing statistically significant temporal variations between consecutive difference images while suppressing stationary sources. This approach enables real-time processing of large image sequences and improves sensitivity to fainter pulsars that may be difficult to identify using traditional source-finding methods.

FITrig produces a \texttt{tLISI} cube, in which two dimensions represent the spatial tile grid of the image and the third dimension corresponds to time. Each element of the cube is a \texttt{tLISI} value ranging from 0 (completely dissimilar) to 1 (identical), quantifying the similarity of the corresponding tile between adjacent difference images. Tiles containing a pulsar exhibit temporal variations that are reflected in their \texttt{tLISI} time series. A Fourier transform is then applied along the temporal axis of the \texttt{tLISI} cube. Harmonic summing is performed on the resulting power spectrum for each spatial tile, producing a two-dimensional (2-D) detection-score matrix on the tile grid. The score matrix is converted to a significance map and tiles whose scores exceed a $5\sigma$-threshold are identified as candidate detections. Connected significant tiles are merged and converted into bounding boxes in the full-resolution image coordinates.

The pulsar position is then estimated within each bounding box in a high-accuracy full-resolution difference image that contains the most significant pulsar signal along the time axis. This difference image is automatically selected by analysing the \texttt{tLISI} cube. The position estimation is performed by subtracting the local median background, selecting locally significant source pixels, and computing an intensity-weighted centroid for the connected component containing the brightest pixel. Finally, the pixel coordinates of the centroids are transformed into Right Ascension and Declination (RA/Dec) using the FITS header keywords. These RA/Dec coordinates define the detected pulsar positions.

\section{Transient-Oriented Imager (TOI)}
\label{FIPsec53}

The design of TOI is motivated by the geometric structure of the visibilities within a single time sample. Within short integration times, the sampled baselines exhibit approximate coplanarity in the frequency domain. This indicates that, for the transient localisation use case considered in this article, conventional full \textit{w}-term correction such as \textit{w}-stacking can be replaced by a reduced-dimensional treatment aligned with the visibility plane for higher efficiency, with a residual correction applied when the departures from coplanarity are not negligible.

\begin{figure*}
\centering
    \begin{subfigure}{0.33\textwidth}\includegraphics[width=\linewidth]{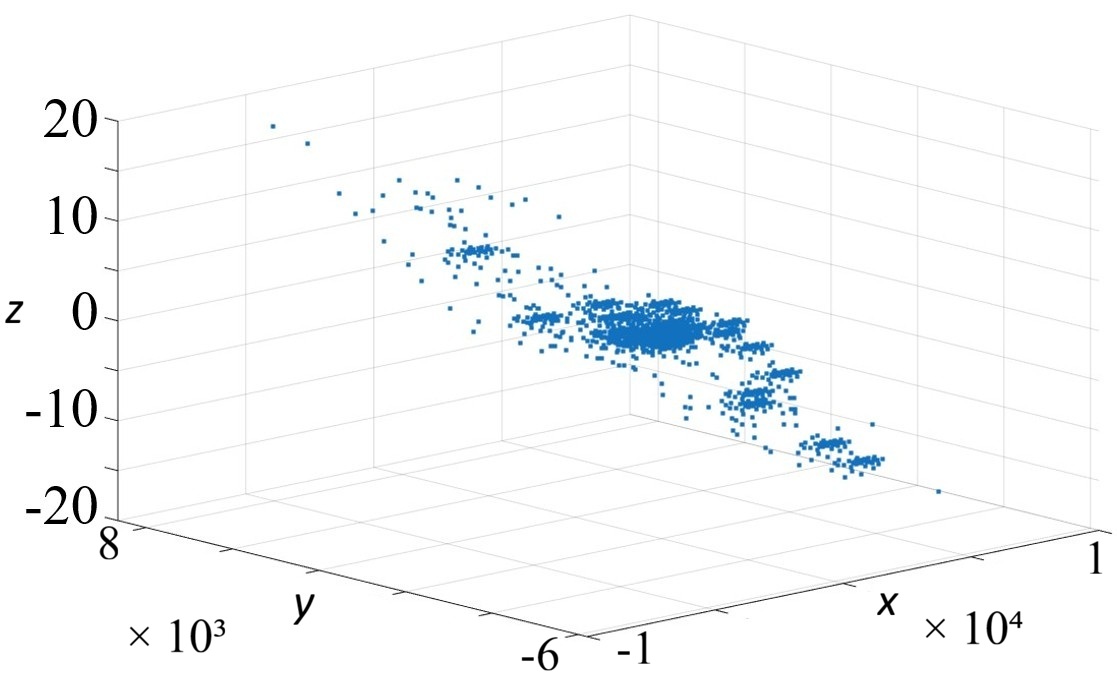}\subcaption{}\end{subfigure}
    \begin{subfigure}{0.33\textwidth}\includegraphics[width=\linewidth]{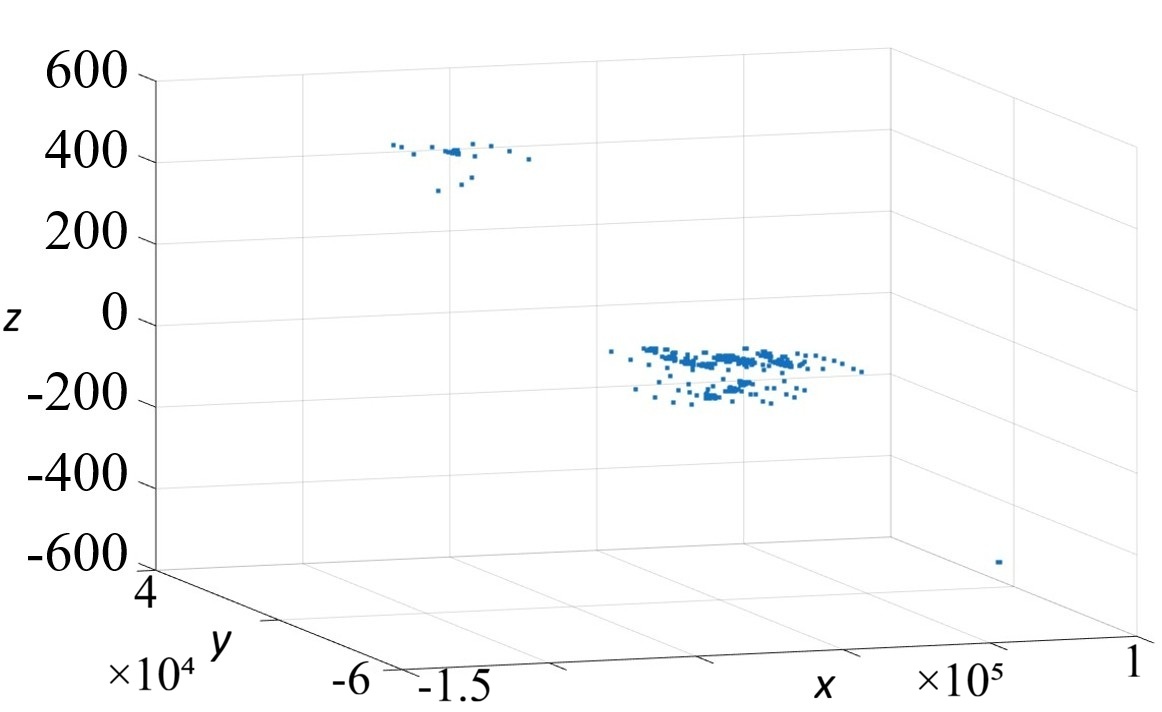}\subcaption{}\end{subfigure}
    \begin{subfigure}{0.33\textwidth}\includegraphics[width=\linewidth]{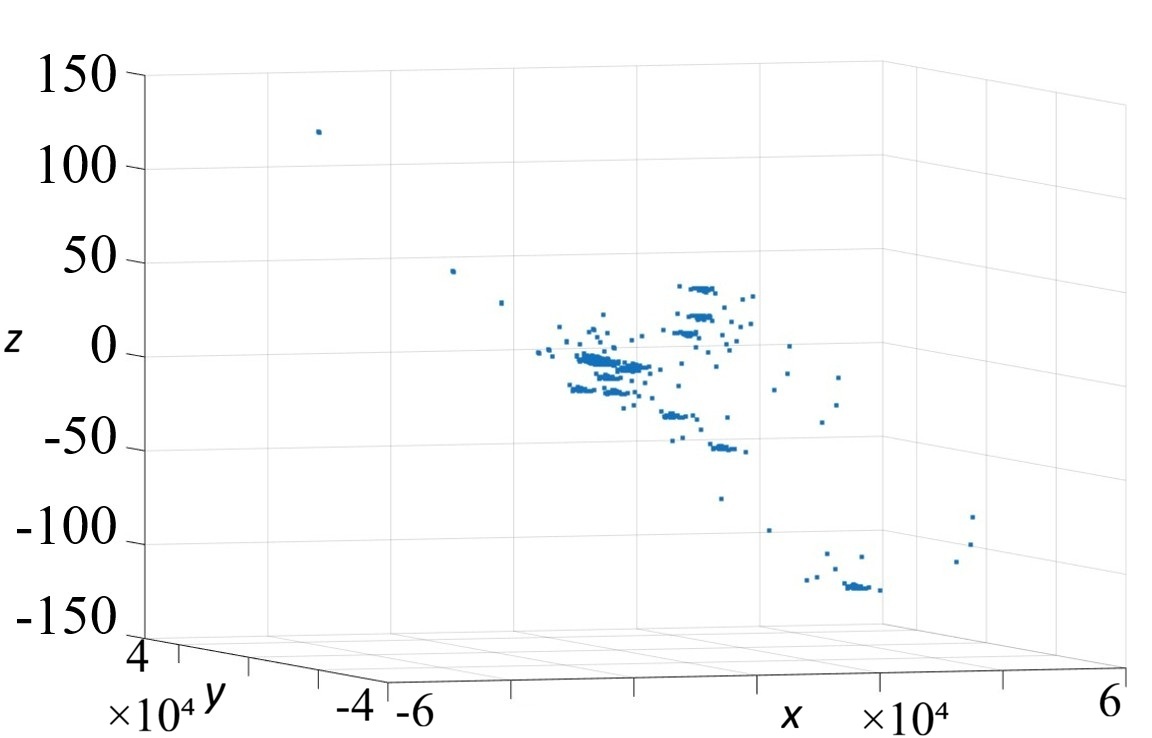}\subcaption{}\end{subfigure}
    \caption{Rigidly rotated baseline distributions for single snapshots simulated from the telescope layouts of (a) MeerKAT, (b) SKA AA2 (MID), and (c) SKA AA2 (MID) with the single antenna located far from the core excluded. The dump time is 2 seconds for MeerKAT and 0.125 seconds for SKA AA2 \citet{skaaa2}. For each snapshot, baselines whose radial distances from the baseline centroid are within 25\% of the maximum radial distance are selected to define a central reference plane. The complete baseline distribution is then rigidly rotated about its centroid so that the central reference plane is aligned with the horizontal plane. The plotted z-coordinate, in metres, therefore represents the signal perpendicular deviation of each baseline from the central reference plane, providing a direct measure of the degree of non-coplanarity.
\label{nearcoplanar}}
\end{figure*}

While this coplanarity holds well for current arrays (e.g., MeerKAT; Fig. \ref{nearcoplanar} (a)), next-generation instruments with longer baselines (e.g., SKA) exhibit small but non-negligible departures from a strictly planar geometry even within a single time sample (Fig. \ref{nearcoplanar} (b)). Here, ``small'' should be understood relative to the baseline scale of the array. Using SKA AA2 as an example, these deviations arise from its extended baseline distribution. Importantly, removing the most distant antenna reduces but does not eliminate the departures (Fig. \ref{nearcoplanar} (c)). Thus, while a full \textit{w}-term correction is not required for the reduced-dimensional transient-imaging strategy used by TOI, a correction beyond a purely planar approximation is still required to preserve imaging accuracy for faint transients measured by next-generation telescopes. It is also worth noting that Fig. \ref{nearcoplanar} (b) shows only AA2; the full SKA-MID configuration, including AA2, AA*, and AA4 \citet{skaarray}, will be even larger, with longer baselines and more significant non-coplanar features within a single snapshot. In such settings, axis-dependent least-square fitting, such as that used in existing snapshot imaging, can produce biased estimates due to its coplanar assumption.

TOI addresses this challenge by applying a Singular Value Decomposition (SVD)-based method directly to the natural three-dimensional (3-D) coordinates derived from the vector-oriented formulation of the visibility equation \citet{RAbook}. Rather than serving only as a replacement for plane fitting, the SVD provides a symmetric, coordinate-independent characterisation of the dominant 2-D subspace in which the baselines reside, while naturally keeping the full 3-D geometry within the measurement equation. Unlike conventional \textit{w}-snapshot approaches \citet{w4,wssnap}, which require additional projection or stacking steps to handle fully 3-D geometries, the SVD formulation incorporates the third dimension directly into the imaging formalism. This enables a coordinate transformation aligned with the array geometry, making the method suitable for real-time imaging and scalable to large interferometric arrays, from approximately coplanar configurations such as MeerKAT to more strongly non-coplanar (described as ``near coplanar'' herein) layouts such as the SKA, see Sections \ref{section32} and \ref{sec34}, respectively, for details.

Crucially, the mathematical framework developed in this section forms the foundation for a GPU-accelerated implementation using Compute Unified Device Architecture (CUDA; \citealt{CUDA}), enabling robust and scalable processing of large-scale datasets (see Section \ref{FIP54}).

Mathematically, the vector-oriented formulation \citet{RAbook} is shown by
\begin{equation}
\mathit{Vis}(\mathbf{b}_\lambda) = \int{A(\sigma)\mathcal{B}_\nu(\sigma)e^{-i2\pi \mathbf{b}_\lambda\cdot\sigma}\mathrm{d}\Omega},
\label{predcr}
\end{equation}
where $\mathbf{b}_\lambda$ represents the baseline vector in units of wavelength $\lambda$, $\sigma$ indicates the source's offset vector representing its angular deviation from the phase centre, $A(\sigma)$ is the primary beam as a function of offset, $\mathcal{B}_\nu$ denotes the sky brightness at frequency $\nu$, and $\mathrm{d}\Omega$ represents the solid-angle element. In practice, the implementation applies primary-beam correction after image reconstruction; for clarity, the following derivation absorbs the primary-beam response into the apparent sky brightness. Equation (\ref{predcr}) illustrates that baselines can be expressed in any coordinate system to define visibilities, where $u,v,w$-coordinate system is one of the choices. 

TOI operates on standard Measurement Sets, maximising accessibility for the astronomical community by eliminating the need for custom data conversions. The UVW coordinates are converted from metres to wavelengths for each frequency channel before imaging. Unless stated otherwise, the derivation below uses one visibility sample indexed by $q$, where $q$ runs over all selected baseline-frequency samples (each sample is a baseline corresponding to a certain frequency, where the unit of the sample is wavelength). Flagging, weighting, polarisation averaging, and multi-frequency synthesis are described in Section \ref{sec35}.

\subsection{Coordinate transformation in the frequency domain}
\label{sec31}

Let $\mathbf{B}=[\mathbf{u},\mathbf{v},\mathbf{w}] \in \mathbb{R}^{N_\mathrm{vis} \times 3}$ denote the baseline matrix for one snapshot, where each row $\mathbf{b}_q ~(q = 1,2,...,N_\mathrm{vis})$ contains one baseline-frequency sample in units of wavelength. For short integration times, these samples lie close to a 2-D subspace in the 3-D UVW coordinate system. TOI uses this property to define a snapshot-dependent coordinate system whose first two axes span the dominant visibility plane.

The basis is estimated from the centred baseline distribution $\tilde{\mathbf{B}}=\mathbf{B}-\bar{\mathbf{B}}$, where $\bar{\mathbf{B}}$ contains the column means of $\mathbf{B}$. Applying SVD, or equivalently principal component analysis (PCA) in this context, to $\tilde{\mathbf{B}}$ gives the dominant orthonormal directions of the snapshot baseline distribution. The first two directions define the visibility plane, and the third direction is chosen to complete a right-handed orthonormal coordinate system. We collect these basis vectors as the rows of the orthogonal transformation matrix $\mathbf{V} \in \mathbb{R}^{3 \times 3}$.

The centring step is used only to estimate the basis. The transformed baseline coordinates $\mathrm{R_1R_2R_3}$ are computed from the original wavelength-scaled baselines, $\mathbf{R} = \mathbf{B}\mathbf{V}^T$, where $\mathbf{R}=[\mathbf{r_1} ~ \mathbf{r_2} ~ \mathbf{r_3}] \in \mathbb{R}^{N_\mathrm{vis} \times 3}$, and $\mathbf{r}_q$ indicates the transformed baseline at row $q$. Thus, the $r_1$- and $r_2$-axes lie on the dominant visibility plane, while $r_3$ measures the residual displacement from that plane. In the 2-D approximation, only $(r_1,r_2)$ are gridded. In the near-coplanar mode (3-D), $r_3$ is retained and corrected as described in Section \ref{sec34}.

\subsection{Coordinate transformation in the spatial domain}
\label{section32}

The spatial-domain coordinates must be transformed consistently with the frequency-domain coordinates. Let 
\begin{equation}
    \mathbf{s}=\begin{bmatrix}
\mathbf{l}\\
\mathbf{m}\\
\mathbf{n}-\mathbf{1}
\end{bmatrix}, ~n=\sqrt{1-l^2-m^2}
\end{equation}
where $(l,m,n)$ are the standard direction cosines relative to the phase centre. We define transformed spatial coordinates
$\mathbf{t} =[
\mathbf{t_1} ~\mathbf{t_2} ~\mathbf{t_3}
]^\mathrm{T}$
by
\begin{equation}
\mathbf{s} = \mathbf{V}^T \mathbf{t}.
\label{equ9}
\end{equation}

Since $\mathbf{V}$ is orthogonal, the Fourier phase is preserved
\begin{equation}
\mathbf{b}_q \cdot \mathbf{s}=\mathbf{r}_q \cdot \mathbf{t} =r_{1,q}t_1+r_{2,q}t_2+r_{3,q}t_3 .
\label{equ10-1}
\end{equation}
For approximate coplanar baselines, $r_{3,q}$ is small over the snapshot. Neglecting the residual coordinate gives 
\begin{equation}
\mathbf{b}_q\cdot \mathbf{s}\simeq r_{1,q}t_1+r_{2,q}t_2 .
\label{equ10}
\end{equation}

The dirty image $I_D$ in the transformed coordinate system can therefore be computed using 2-D Fourier reconstruction (superscript $2$ for coplanar usage of $\mathbf{R}$),
\begin{equation}
    I_D^{(2)}(t_1,t_2)=\frac{1}{\omega N_\mathrm{vis}}\sum^{N_\mathrm{vis}}_{q=1}\mathrm{Vis}_q \mathrm{exp}[i2\pi(r_{1,q}t_1+r_{2,q}t_2)],
\end{equation}
where $\mathrm{Vis}_q$ is the visibility value corresponding to the baseline row $q$ after flagging, weighting, polarisation selection, and frequency handling. The factor 
\begin{equation}
\omega=\left|\frac{\partial (l,m)}{\partial (t_1,t_2)}\right|=|V_{11}V_{22}-V_{21}V_{12}|
    \label{equ11}
\end{equation}
accounts for the area change between the standard $(l,m)$ coordinates and the transformed $(t_1,t_2)$ coordinates.

The image produced by this 2-D reconstruction lies on the transformed $t_1,t_2$-plane. It is mapped back to the standard image grid using the re-projection process described in Section \ref{sec33}. The 2-D approximation is appropriate when the residual phase $\mathrm{exp}(2\pi r_{3,q}t_3)$ is negligible over the valid image region. When this condition is not satisfied, TOI uses the near-coplanar correction described in Section \ref{sec34}.

\subsection{Re-projection process}
\label{sec33}

The dirty image produced by the Fourier reconstruction is sampled on the transformed $t_1,t_2$-plane. Since this plane is tilted relative to the standard $l,m$-plane, the image must be re-projected before it is passed to the downstream transient localisation stage.

This re-projection is based on the geometry of the SIN projection in the FITS World Coordinate System (WCS) \citet{wcs} which is suitable for aperture synthesis in radio astronomy; however, TOI does not evaluate the standard WCS equations directly. The conventional calculation proceeds through native spherical coordinates $(\phi,\theta)$; for comparison, it is summarised in Appendix \ref{appendix1}. In the current TOI implementation, we instead use an algebraically reduced pixel-to-pixel mapping that is more suitable for GPU execution.

The process begins by mapping pixel coordinates $(p_1,p_2)$ on the tilted plane to angular coordinates $(x,y)$: 
\begin{equation}
\left\{ 
\begin{matrix}
x = -\beta\left( p_1 - \left(\frac{N}{2}+1\right)\right)\\
y = \beta\left(p_2-\left(\frac{N}{2}+1\right)\right) ~~\\
\end{matrix}
\right . ,
\end{equation}
where $\beta$ is the cell size in degrees and the image size is $N \times N$.

To simplify the trigonometry, equivalently, we let
\begin{equation}
    h=\sqrt{x^2+y^2}, \qquad r_0=\frac{180}{\pi}.
\end{equation}
The valid SIN projection domain is $h\leq r_0$. For valid pixels, define
\begin{equation}
    \hat{x}=\left\{\begin{matrix}
         x/h, \qquad h\ne0\\
         0,\qquad~~~~ h=0
    \end{matrix}\right.
    \qquad
    \hat{y}=\left\{ \begin{matrix}
        -y/h, \qquad h\ne 0\\
        1, \qquad ~~~~~~h=0
    \end{matrix}\right .
\end{equation}
The radial SIN correction is 
\begin{equation}
    z=1-\sqrt{1-(h/r_0)^2}.
\end{equation}
Near the projection centre, we use the algebraically equivalent form
\begin{equation}
      z =
      \frac{(h/r_0)^2}
      {1+\sqrt{1-(h/r_0)^2}},
  \end{equation}
which avoids loss of precision when $h/r_0$ is small.

The tilt of the transformed image plane is determined by the SVD-derived coordinate transformation matrix $\mathbf{V}$. Define
\begin{equation}
        \xi = \frac{V_{31}}{V_{33}}, \qquad
        \eta = \frac{V_{32}}{V_{33}}.
\end{equation}
The reduced direct mapping to the standard image plane is then
\begin{equation}
    \left\{\begin{matrix}
        x' = h\hat{x} + r_0 z \xi\\
      y' = h\hat{y} + r_0 z \eta 
    \end{matrix}\right. .
\end{equation}
Finally, these angular coordinates are converted to output pixel coordinates,
\begin{equation}
    \left\{ \begin{matrix}
        p'_1 = -\frac{x'}{\beta}+\frac{N}{2}+1\\
      p'_2 =  \frac{y'}{\beta}+\frac{N}{2}+1~~~
    \end{matrix}\right. .
\end{equation}

This direct mapping is algebraically equivalent to the SIN geometry used by the conventional WCS formulation, but it is not evaluated through sequential WCS coordinate conversions. Existing WCS implementations are predominantly CPU-based and are not designed for large-scale, highly parallel processing. In contrast, TOI uses the reduced form above so that each transformed-plane pixel can be re-projected independently, enabling scalable, reproducible, and high-performance image re-projection.

\subsection{Near coplanar scenario}
\label{sec34}

For longer-baseline arrays, the residual coordinate $r_3$ can be small relative to the in-plane coordinates but still large enough to produce a non-negligible phase error. The removal of this term can smear faint sources and reduce image fidelity. In this case, TOI retains the third coordinate while continuing to use 2-D FFTs over the dominant $r_1,r_2$-plane.

For a given image-domain coordinate $(t_1,t_2)$, the corresponding value of $t_3$ is constrained by the celestial sphere. Solving this constraint gives
\begin{equation}
    t_3(t_1,t_2)=-V_{33} + \mathrm{sgn}(V_{33}) \sqrt{1-(t_1+V_{13})^2-(t_2+V_{23})^2} .
    \label{eqnt3}
\end{equation}
The branch is selected so that $t_3$ remains close to zero near the imaging plane. A complete derivation is available in Appendix \ref{appB1}. 

Retaining the residual phase (superscript $3$ for 3-D $\mathbf{R}$ coordinates) gives the near-coplanar dirty image
\begin{equation}
    I_D^{(3)}(t_1,t_2)=\frac{1}{\omega N_\mathrm{vis}}\sum^{N_\mathrm{vis}}_{q=1}\mathrm{Vis}_q \mathrm{exp}[i2\pi(r_{1,q}t_1+r_{2,q}t_2+r_{3,q}t_3)].
    \label{eqn10}
\end{equation}

To compute it efficiently, the implementation divides the range of $r_3$ into slabs. For slab $p$, let $r_{3,p}$ be the slab centre and $\Delta r_3$ be its half-width. For each visibility (indexed $q$) in this slab (indexed $p$), define 
\begin{equation}
    s_q = \frac{r_{3,q}-r_{3,p}}{\Delta r_3}, ~ -1 \leq s_q \leq 1.
\end{equation}

The residual phase then separates as
\begin{equation}
    e^{i2\pi r_{3,q}t_3}=e^{i2\pi r_{3,p}t_3}e^{i\alpha s_q}, ~\text{where} ~\alpha = 2 \pi \Delta r_3 t_3 .
    \label{eqn12}
\end{equation}

The second factor can be expanded using Chebyshev polynomials through the Jacobi-Anger identity,
\begin{equation}
    e^{i\alpha s_q}\simeq \sum_{k=0}^{K-1}c_k(\alpha)T_k(s_q),
\end{equation}
where 
\begin{equation}
    \begin{cases}
        c_0(\alpha)=J_0(\alpha)\\
    c_k(\alpha)=2 i^k J_k(\alpha), k>0
    \end{cases} .
\end{equation}
Here, $J_k$ is the Bessel function of the first kind, $T_k$ is the Chebyshev polynomial of degree $k$, and $K$ is the Chebyshev truncation order. The derivation is given in Appendix \ref{appB2}. 

Substituting this expansion into Equation (\ref{eqn10}) gives
\begin{equation}
    I_D^{(3)}(t_1,t_2)\simeq \frac{1}{\omega N_\mathrm{vis}}\sum_p e^{i2\pi r_{3,p}t_3}\sum_{k=0}^{K-1}c_k(2 \pi \Delta r_3 t_3)M_{p,k}(t_1,t_2),
\end{equation}
where the Chebyshev moment for slab $p$ and order $k$ is
\begin{equation}
M_{p,k}(t_1,t_2)=\sum _{q\in S_p}T_k(s_q)\mathrm{Vis}_q e^{i2\pi(r_{1,q}t_1+r_{2,q}t_2)} .
\end{equation}
Here, $S_p$ is the set of visibility samples assigned to slab $p$. Each $M_{p,k}$ is computed using a 2-D gridding operation followed by an inverse FFT on the $(r_1,r_2)$ grid. The final image is formed by summation over slabs and Chebyshev orders.

The number of slabs and Chebyshev terms are selected from the maximum residual phase in the valid image region. Let
\begin{equation}
    t_{3,\mathrm{max}}=\max_{(t_1,t_2)\in\mathcal{D}}|t_3(t_1,t_2)|,
\end{equation}
where $\mathcal{D}$ is the valid image-domain region. Let $r_{3,\mathrm{min}}$ and $r_{3,\mathrm{max}}$ are the minimum and maximum residual baseline coordinates in the snapshot, the full residual phase scale is
\begin{equation}
  \Phi_{\max} = 2\pi t_{3,\mathrm{max}} \frac{r_{3,\mathrm{max}}-r_{3,\mathrm{min}}}{2}.
  \label{eqn26}
\end{equation}
We choose
\begin{equation}
    N_{\mathrm{slab}}=\max \left(1,\left\lceil \frac{\Phi_{\max}}{\alpha_{\mathrm{target}}}\right\rceil \right),
    \label{eqn27}
\end{equation}
with $\alpha_\mathrm{target}=10$ by default. The Chebyshev truncation order is selected from $K \in \{8,16,24,32\}$ by sampling the complex approximation error over $0\leq \alpha \leq \alpha_\mathrm{slab}$ and $-1 \leq s_q \leq 1$, where $\alpha_\mathrm{slab}=2\pi t_{3,\mathrm{max}}\Delta r_3$. The smallest $K$ whose maximum sampled error $\gamma$ is below $10^{-2}$ is used; otherwise, $K=32$ is used.

In the near-coplanar scenario, TOI requires $N_\mathrm{slab}K$ Chebyshev moment reconstructions, but still avoids a full 3-D Fourier transform.

\subsection{Imaging modes and data handling}
\label{sec35}

TOI supports two imaging modes. The 2-D mode uses the transformed $r_1,r_2$-coordinates and applies a single gridding and inverse FFT operation. This mode is appropriate for snapshots whose baselines are sufficiently coplanar. On the other hand, the near-coplanar mode naturally keeps the residual coordinate $r_3$ and applies the Chebyshev correction described in Section \ref{sec34}.

The input data are read from standard Measurement Set columns. This design ensures that the imager and pipeline remain portable and accessible to a wider range of users. The UVW column provides the baseline coordinates in metres, the DATA column provides the visibilities (complex values), the FLAG column indicates invalid samples, the WEIGHT\_SPECTRUM column provides frequency-dependent weights when available, and CHAN\_FREQ provides the observing frequencies. Before applying the SVD and all subsequent imaging equations, each UVW coordinate is converted from metres to units of wavelength for its corresponding frequency channel.

Multi-frequency synthesis (MFS) is handled by treating each baseline-frequency pair as one visibility sample. Flagged samples are excluded by assigning zero weight. When frequency-dependent weights are available, they are applied before gridding; otherwise, unit weights are used. The implementation averages the selected parallel-hand polarisations using their corresponding weights.

TOI produces dirty snapshots. Deconvolution is not performed because the downstream FITrig localisation stage operates directly on dirty images. When used inside FIP-TOI, the dirty snapshots remain on the GPU and are passed directly to FITrig. When TOI is used as a standalone imager, the dirty snapshot is transferred to the host and written as a FITS image.

\section{GPU acceleration}
\label{FIP54}

The large amount of parallelism available in image processing and the search for transient events in large images make FIP a natural candidate for GPU acceleration. Consequently, FIP leverages the CUDA Toolkit for NVIDIA GPUs to accelerate its critical functions. Numerous elements of its design are tuned to the strengths and quirks of GPUs. FIP is designed to make the best possible use of GPU.

\begin{itemize}
    \item \textbf{Maximising GPU utilisation:} The GPU must always be kept busy with \textit{something} to do. There must always be work waiting for it, as soon as it is done with the previous work. If the GPU can work on multiple things in parallel, that possibility should be made available to it. 
    \item \textbf{Avoiding synchronisation:} The GPU operates by nature asynchronously and downstream of the CPU giving it directions. Work submitted to the GPU is carried out ``eventually''. If the CPU requires the result of an operation, it must await that operation to arrive at the head of its queue and complete, possibly allowing the GPU to become idle afterwards. This is also called a \textit{pipeline flush} and it creates a \textit{pipeline bubble}, a computational void.
    \item \textbf{Minimising usage of limited bandwidth:} Moving data between a GPU's cores and its memory or the rest of the system is fast, but not limitlessly so. It is much more expensive than a floating-point maths operation, in both energy and time. A GPU stalled awaiting data is wasting cycles. Avoiding unnecessary data movement or long waits for data is therefore important.
\end{itemize}

If the GPU is not constantly supplied with a large amount of work, or is frequently required to pause, or must constantly ``spill'' its work to memory and ``reload'', then that GPU will not be well utilised. FIP avoids these pitfalls.

\subsection{Pipelining}
\label{sec41}

FIP is very deeply pipelined. The FIP pipeline contains no less than \textit{\textbf{8}} stages, one for each natural step of processing, from first load to final write-out. These processing stages communicate and transform data through interstage \textit{ring-buffers}. At no time is the GPU idle, short of the CPU failing to supply input data or evacuating output data fast enough. Even data transfer to and from the GPU is overlapped with computation.

\begin{figure*}
    \includegraphics[width=0.75\textwidth]{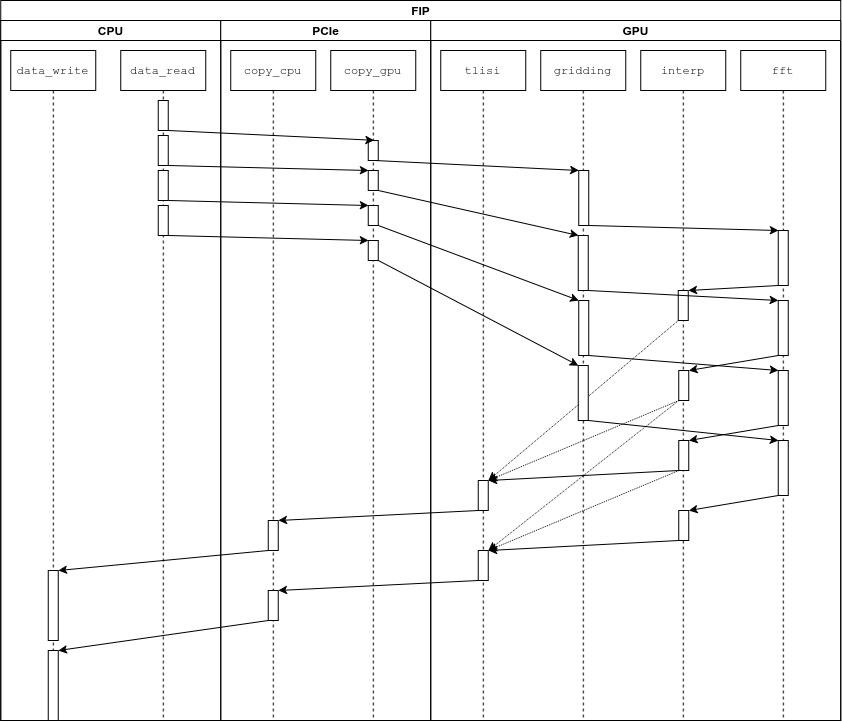}
    \caption{Diagram of FIP-TOI 8-stage CUDA pipeline, for the first 4 snapshots of processing. For the first two snapshots, tLISI cannot yet be evaluated, so processing ends early; From the third snapshot onwards, tLISI can be evaluated and is saved, so the remaining CUDA streams are in use. Each stream is independent and its execution may be concurrent with others. Edges represent the primary interlocking dependencies, as well as the data flow.
    \label{fig-pipeline}}
\end{figure*}

FIP uses \textit{eight} CUDA streams corresponding to each stage. Each is responsible for its titular operation, and hands over its output to the input of the subsequent stage. A sequence diagram displaying this process is shown in Fig. \ref{fig-pipeline}.

\begin{enumerate}
    \item\texttt{data\_read}: Source. Data read from Measurement Set on-disk.
    \item\texttt{copy\_gpu}: Responsible for copying visibilities and coordinates data from the CPU to the GPU.
    \item\texttt{gridding}: Place visibilities on a grid.
    \item\texttt{fft}: Inverse FFT, the most computationally intensive task. Uses single-precision \texttt{cufftExecC2C} function from the CUDA Fast Fourier Transform (cuFFT) library\footnote{\url{https://docs.nvidia.com/cuda/cufft/}}
    \item\texttt{interpolate}: Interpolate a time-domain image.
    \item\texttt{tlisi}: The tLISI transient detector over time-sequential images.
    \item\texttt{copy\_cpu}: Responsible for copying tLISI analyses from the GPU to the CPU.
    \item\texttt{data\_write}: Sink. Write to disc.
\end{enumerate}

At the beginning, the \texttt{data\_read} stream reads the next snapshot to be processed, while at the end the \texttt{data\_write} stream writes out a slice of an analysis data cube into a FITS file. Here, FITS file is merely used as a container data format to store values, rather than images. The FITS format is portable, flexible and well-supported as an input by many software packages. The executions of all these streams overlap, and multiple snapshots can be in flight simultaneously at different stages of processing (as many as eight, due to the eight stages).

\subsection{Removal of synchronisation}

FIP eliminates \textit{all} GPU-to-host synchronisation. Great care is taken to avoid any explicit or implicit synchronisation operation in the main loop. This includes the full removal of all memory allocations within the pipeline\footnote{\texttt{cudaMalloc()} synchronises in order to modify the GPU's memory map.}. Furthermore, copies of data to and from the GPU exclusively use pinned-memory ring-buffers (\texttt{cudaMallocHost()}) and asynchronous transfers (\texttt{cudaMemcpyAsync()}), as only transfers to such buffers can be non-blocking\footnote{CUDA transfers to non-pinned memory must allocate a temporary \textit{bounce buffer} and thus must obligatorily synchronise}.

\subsection{Inter-stage Synchronisation}

Extensive use is made of two sets of CUDA ``event'' primitives to provide the interlocking required between pipeline stages without requiring the CPU to become an active participant in this interlocking.

\subsubsection{Events}

Pipeline stages must still synchronise with each other to perform safe handovers of data between stages, and prevent asynchronously-launched GPU kernels from consuming stale data or overwriting live data. However, these synchronisations do not require the involvement of the host CPU. We heavily rely on a feature of the CUDA Toolkit's API called \textit{events} (\texttt{cudaEvent\_t}) to asynchronously establish dependencies between CUDA streams. CUDA events are lightweight markers of progress. They may be enqueued onto a stream such as \texttt{gridding} using \texttt{cudaEventRecord(evt, gridding)}. A dependency on that event can simultaneously be enqueued onto a different stream, such as \texttt{fft}, using \texttt{cudaStreamWaitEvent(fft, evt)}. When the GPU asynchronously reaches that event in the first queue (\texttt{gridding}), the event is considered \textit{complete}, and the other queue (\texttt{fft}) is unblocked and allowed to proceed.

\subsubsection{Ring buffering}

A \textit{ring buffer} is a linear buffer treated as a circle, with wraparound from the last element to the first. Data added to it advances one cursor, and data removed from it advances another, modulo the number of elements in the buffer. The number of elements that can be stored in a ring buffer may be referred to as its \textit{depth}. The ring buffer supports various single-producer single-consumer scenarios.

The most trivial example of a ring buffer is the double buffer; It has depth 2. Double-buffering is a technique that allows a producer to write data into a ``front'' buffer, while a consumer is using an intact copy of previous data in the ``back'' buffer. FIP uses depth-2 (double) buffers for copying data in and out of the GPU.

The principle generalises to depth 3 (triple buffering) and to greater depths. Greater depth achieves greater decoupling between stages and can better absorb temporary slowdowns in other parts of the pipeline. However, greater depth also increases memory consumption, leading to a classic time-memory trade-off. And sometimes, greater depth is an algorithmic necessity.

FIP relies on a depth-3 ring-buffer (triple buffer) for the gridding data, because FFT is executed over them in-place. Because the FFT is executed in-place, safety requires that either the gridding stage or the interpolation stage would have to block for the FFT to complete. This is unacceptable for a high-performance pipeline. With a triple buffer, no stage is required to stall. The \texttt{interpolation} stage could read from buffer \texttt{[0]}, the \texttt{fft} stage could operate in-place on buffer \texttt{[1]}, and the \texttt{gridding} stage could write into buffer \texttt{[2]}, all concurrently and safely.

FIP also relies, uniquely, on a depth-5 ring-buffer for the images. The tLISI triggering algorithm requires the past three images for analysis; therefore, images must be preserved for two more iterations of the pipeline than the grids. Keeping images on the GPU for on-GPU triggering also avoids transferring the images out of the GPU and straining the GPU's limited I/O bandwidth to the outside world.

\subsection{Fusion}

FIP employs fused operations whenever possible, combining as many related computations in one GPU kernel as reasonably feasible. Fused operations avoid ``spill and reload'', the unnecessary round-trips through memory of data that could instead be immediately used for the next stage of processing. Such unnecessary traffic grinds away at memory bandwidth for other kernels and reduces computational intensity, increasing the likelihood of GPU idle cycles. Furthermore, the fusion process occasionally exposed optimisation and simplification opportunities at a mathematical, numerical, or algorithmic level. Several such opportunities were noticed and exploited in the construction of FIP.

Notably, the conversion between coordinate systems was rich in optimisation opportunities. TOI utilises custom-developed functions written according to the mathematical principles detailed in Section \ref{FIPsec53}. As a result of simplifications made available by fusion, numerous trigonometric function calls like \texttt{sin()}, \texttt{cos()}, \texttt{asin()},  \texttt{acos()} and \texttt{atan2()} were entirely eliminated or transformed through identities to the cheaper \texttt{sqrt()} for which hardware support exists. Finally, by eliminating dependencies on external, CPU-only libraries like Nifty Gridder \citet{nifty} and WCSLIB \citet{wcs,wcs1}, this design enhances the pipeline's flexibility and portability, and further avoids synchronisation points or CPU-GPU traffic.

\subsection{Output}

The data downloaded from the GPU is the output of the transient finding stage of FIP, FITrig. In it, images are divided into multiple tiles, and the rate of temporal change for each tile is determined by comparing the three youngest dirty snapshots from the ring buffer to localise potential transient events. A tLISI value is output for each tile in one comparison between the three snapshots. The lower the tLISI value, the more likely the tile contains transient events. For each slice, its tLISI matrix is materialised on the GPU in a depth-2 ring buffer. Because it will shortly be overwritten, these matrices are quickly evacuated by device-to-host asynchronous copies to the CPU, then appended to a FITS file. Stacking these matrices along the time axis constructs a cube. Accelerated by GPUs, the functional diagram of the tLISI kernel is shown in Fig. \ref{functlisi}. A comprehensive presentation of tLISI and FITrig is presented in \citealt{FItrigger}.

The output tLISI values for each snapshot are saved slice-by-slice in a FITS output file, ready for refined localisation to obtain RA/Dec of transients, as described in Section \ref{FIPsec52}. As these slices are orders of magnitude smaller than the input data and are fully overlapped with data input and processing, they add negligible time to the pipeline.

\section{Performance}
\label{FIP55}

This section presents experiments designed to assess the performance of FIP-TOI (TOI + FITrig) compared to WSClean-based FIP (WSClean + FITrig) and SKA-SDP-FIP. 
 
In this section and in the next (Section \ref{FIP553}), we use real datasets of PSR J0901-4046 \citet{realpul2} and PSR J1703-4902 \citet{psr1,psr2}, both observed by MeerKAT. Besides, we use simulated datasets generated by PulSKASim \citet{pulskasim}, a pulsar simulator for SKA-scale interferometric observations, with the support of Oxford's Square Kilometre Array Radio-telescope simulator (OSKAR; \citealt{OSKAR1}). We use telescope configurations of SKA1-LOW, SKA AA2 (from SKA1-MID), and MeerKAT in the simulations.

\subsection{Performance of imagers}
\label{FIP551}

We use simulated datasets to evaluate the performance of TOI against other imagers, with the FOV set to a wide field of 3 degrees. Within this FOV, 16 sources are arranged in a $4 \times 4$ grid, evenly distributed with each spacing in both horizontal and vertical directions. Among these 16 sources, one of them intentionally shifts 0.2 degrees to serve as the direction calibrator. This distribution spans the entire image, enabling an impartial assessment of the imager's performance across all image regions. The same SBD is simulated using various telescope configurations to avoid bias. Here, we use the layouts of SKA1-LOW, SKA AA2 (MID), and MeerKAT telescopes. In the experiment, SKA1-LOW functions at a lower frequency of 60 MHz, whereas both SKA AA2 (from SKA1-MID) and MeerKAT function at a higher frequency of 1.2 GHz. The details of the simulation parameters are listed in Table \ref{simupara}. 

\begin{figure*}
\centering
    \begin{subfigure}{0.18\textwidth}\includegraphics[width=\linewidth]{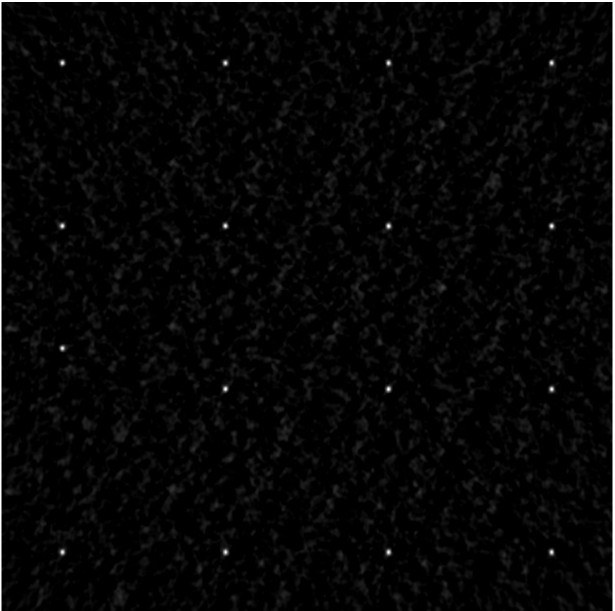}\subcaption{}\end{subfigure}
    \begin{subfigure}{0.18\textwidth}\includegraphics[width=\linewidth]{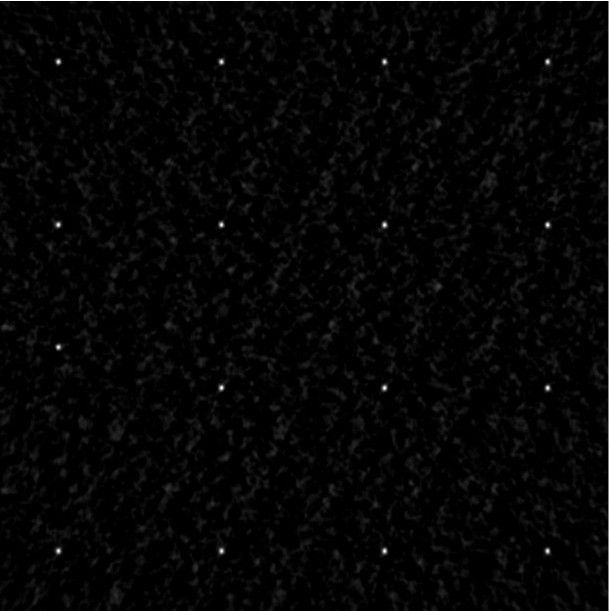}\subcaption{}\end{subfigure}
    \begin{subfigure}{0.18\textwidth}\includegraphics[width=\linewidth]{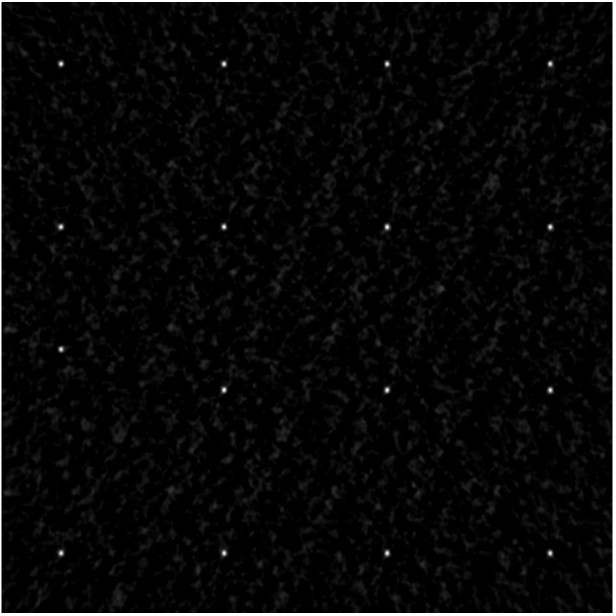}\subcaption{}\end{subfigure}
    \begin{subfigure}{0.18\textwidth}\includegraphics[width=\linewidth]{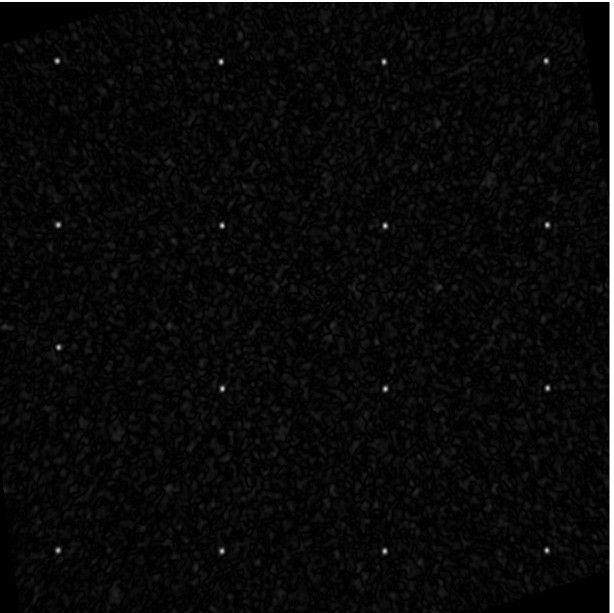}\subcaption{}\end{subfigure}
    \begin{subfigure}{0.215\textwidth}\includegraphics[width=\linewidth]{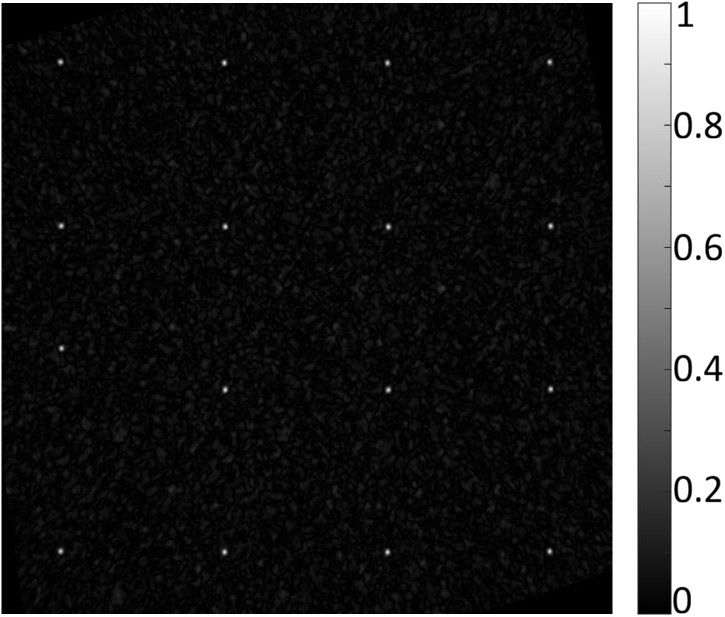}\subcaption{}\end{subfigure}
    \begin{subfigure}{0.18\textwidth}\includegraphics[width=\linewidth]{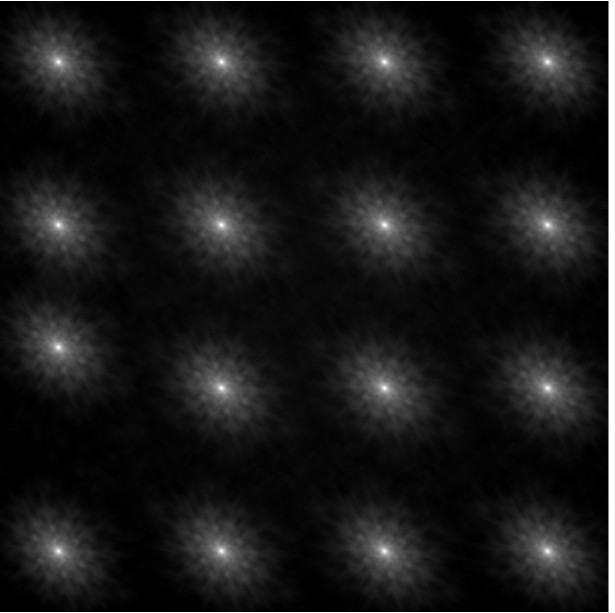}\subcaption{}\end{subfigure}
    \begin{subfigure}{0.18\textwidth}\includegraphics[width=\linewidth]{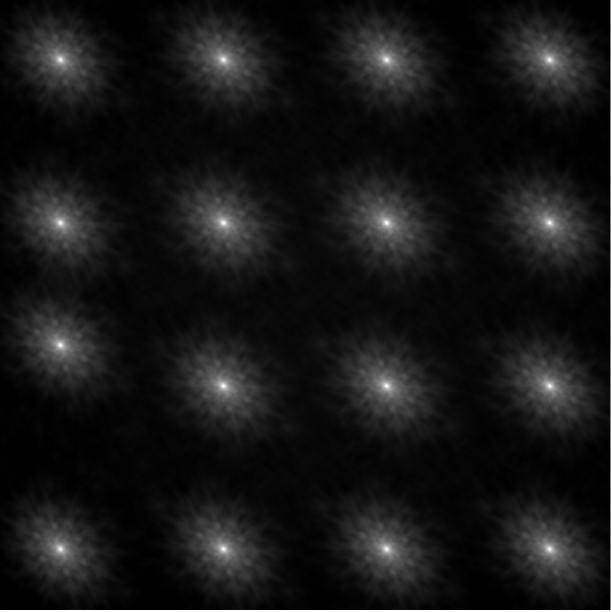}\subcaption{}\end{subfigure}
    \begin{subfigure}{0.18\textwidth}\includegraphics[width=\linewidth]{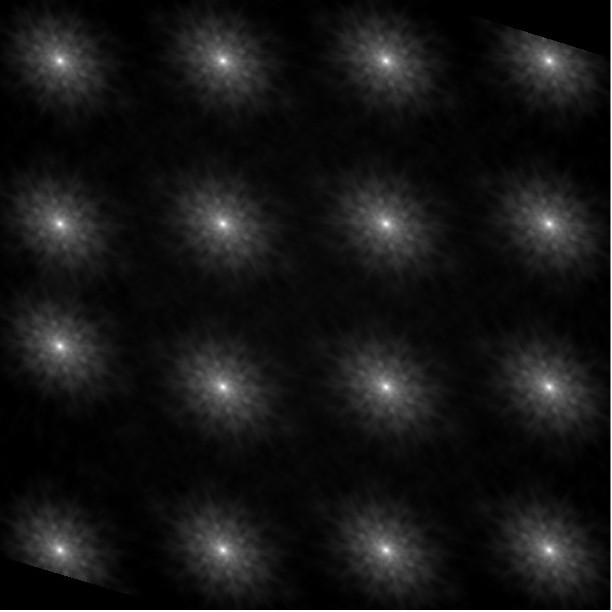}\subcaption{}\end{subfigure}
    \begin{subfigure}{0.18\textwidth}\includegraphics[width=\linewidth]{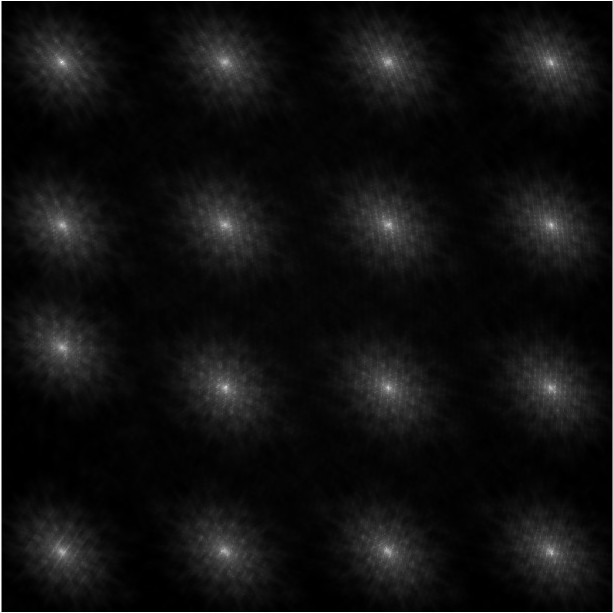}\subcaption{}\end{subfigure}
    \begin{subfigure}{0.215\textwidth}\includegraphics[width=\linewidth]{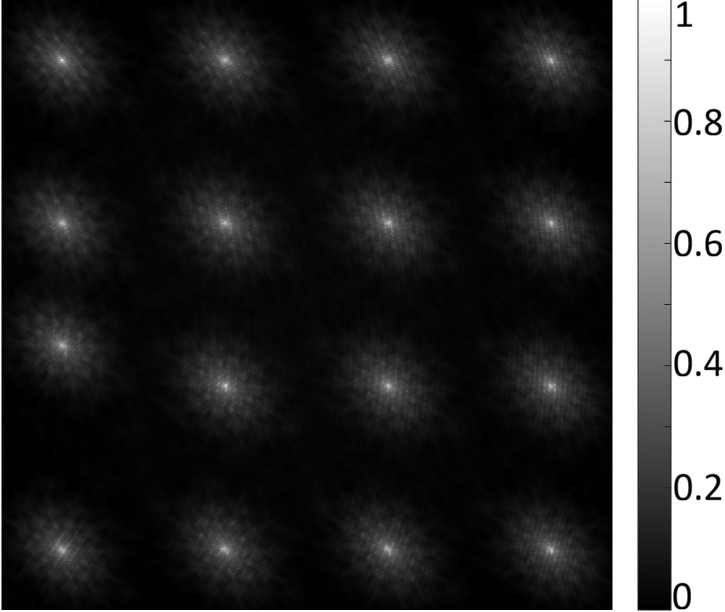}\subcaption{}\end{subfigure}
    \begin{subfigure}{0.18\textwidth}\includegraphics[width=\linewidth]{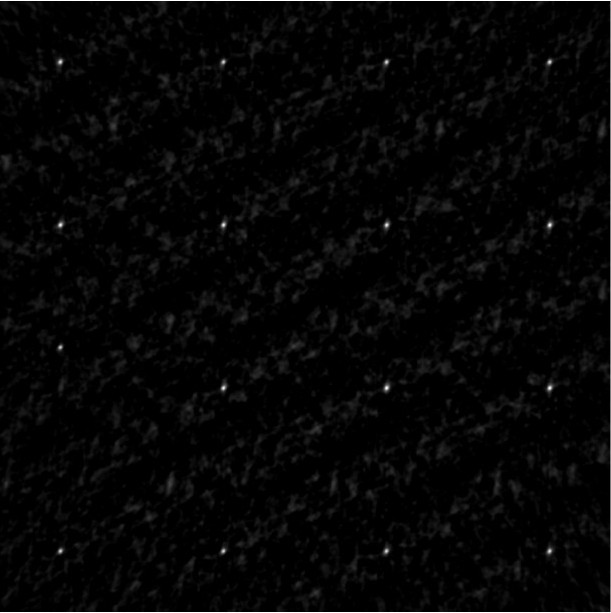}\subcaption{}\end{subfigure}
    \begin{subfigure}{0.182\textwidth}\includegraphics[width=\linewidth]{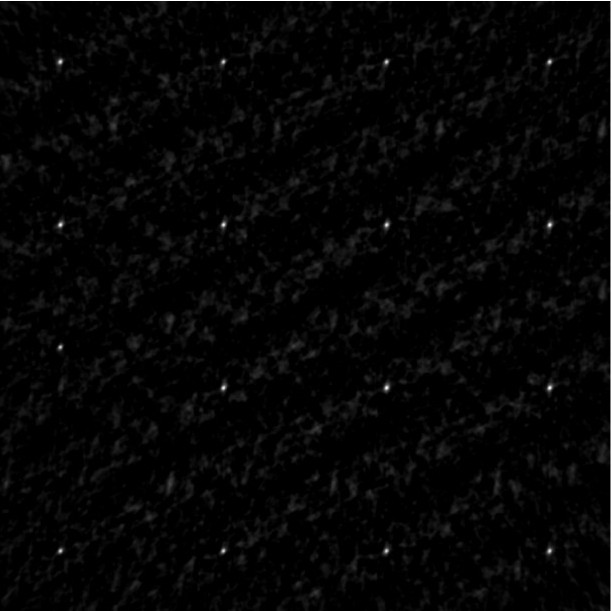}\subcaption{}\end{subfigure}
    \begin{subfigure}{0.182\textwidth}\includegraphics[width=\linewidth]{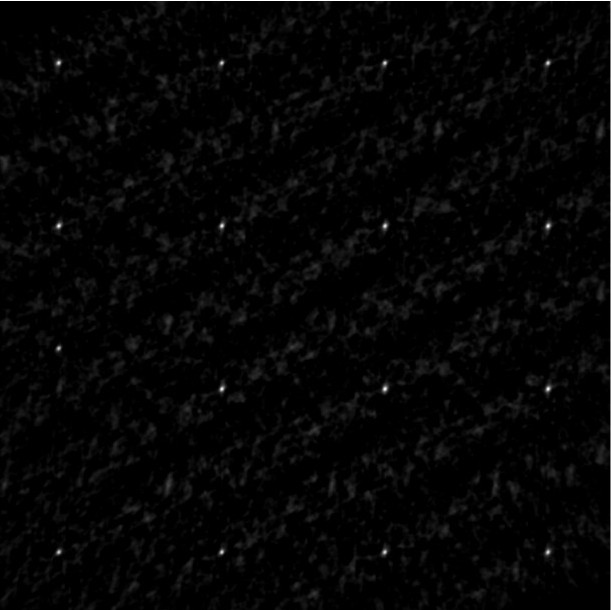}\subcaption{}\end{subfigure}
    \begin{subfigure}{0.182\textwidth}\includegraphics[width=\linewidth]{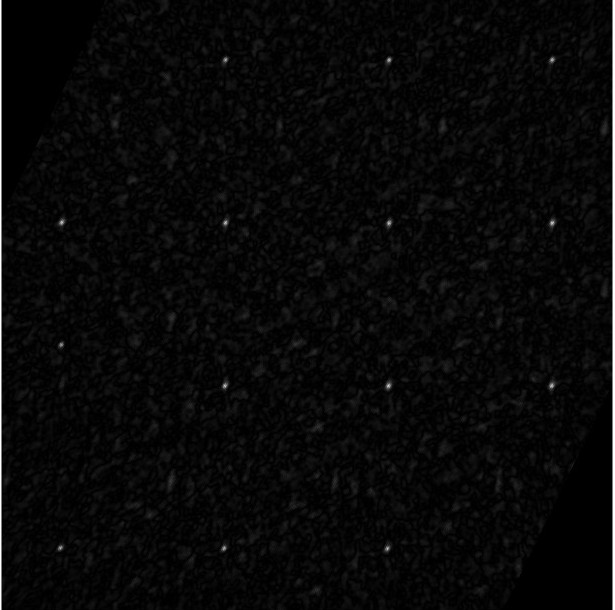}\subcaption{}\end{subfigure}
    \begin{subfigure}{0.215\textwidth}\includegraphics[width=\linewidth]{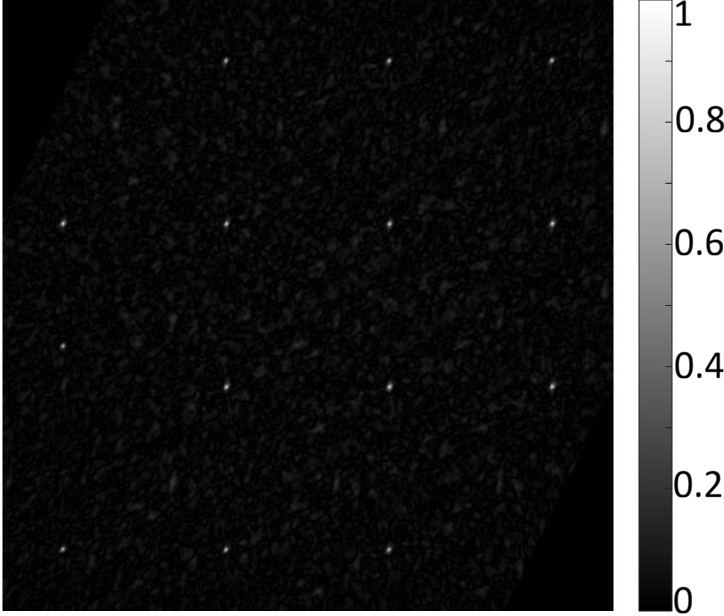}\subcaption{}\end{subfigure}
    \caption{Dirty snapshots for (top row) MeerKAT, (middle row) SKA1-LOW, and (bottom row) SKA AA2 with different imaging algorithms shown in columns 1-5: WSClean (w-stacking), SKA-SDP (w-stacking), WSClean (w-snapshot), TOI (2-D), and TOI (3-D). The images are shown with comparable intensity scales for illustration purpose.
\label{imgqacomp1}}
\end{figure*}

The dirty snapshots produced by the WSClean, TOI, and SKA-SDP imager are illustrated in Fig. \ref{imgqacomp1}. These images have dimensions of $4096 \times 4096$ pixels, with each pixel corresponding to a cell size of 0.000732 degrees. Natural weighting mode is chosen to preserve visibility weights and maximise sensitivity. Single-precision floating-point format (FP32) is used in the imaging. In this experiment, five imaging algorithms are run. SKA-SDP imager uses \textit{w}-stacking for the \textit{w}-correction. WSClean is run with two modes separately: the default w-stacking mode and the w-snapshot mode, where, when running w-snapshot, a subsequent re-projection is conducted by \texttt{astropy} and \texttt{reproject} libraries. TOI runs 2-D and 3-D versions, corresponding to the methods illustrated in Sections \ref{section32} and \ref{sec34}. The effective region beyond the missing corners of snapshot-based imaging results corresponds to the overlap between the transformed image grid and the final reprojected image grid, which is at least 91.42\% of the entire image \citet{xiaotongthesis}.

The parameters used to control the Chebyshev truncation in TOI (3-D) are illustrated in Table \ref{cheby}. These parameters are defined in Equations (\ref{eqn26}) and (\ref{eqn27}). Higher values of $\Phi_{\max}$ indicate more significant non-coplanar features. These values are selected by the algorithm automatically following the standard described in Section \ref{sec34}.

\begin{table}
\centering
\caption{Chebyshev correction parameters used in TOI (3-D), with $\alpha_\mathrm{target}=10$ and $\gamma=0.01$.\label{cheby}}
\begin{tabular}{cccccc}
 \hline
Telescope & $\Phi_{\max}$ & $N_\mathrm{slab}$ & $\alpha_\mathrm{slab}$ & $K$ & $N_\mathrm{slab}K$\\
 \hline
MeerKAT & 5.11 & 1 & 5.11 & 16 & 16\\
SKA1-LOW & 6.34 & 1 & 6.34 & 16 & 16\\
SKA AA2 & 163.63 & 17 & 9.63 & 16 & 272\\
\hline
\end{tabular}
\end{table}

The results of the analysis of Fig. \ref{imgqacomp1} are illustrated in Table \ref{table51}. The position error is calculated as the Euclidean distance between the source positions produced by each imager and those obtained using WSClean (\textit{w}-stacking) in the RA/Dec coordinate space; this is a fair reference because WSClean is widely used by the astronomical community.

\begin{table*}
\centering
\caption{Performance of imagers.\label{table51}}
\begin{tabular}{ccccccc}
 \hline
Telescope & Metric & WSClean (stack) & SKA-SDP (stack) & WSClean (snap) & TOI (2-D) & TOI (3-D)\\
 \hline
MeerKAT & Position error (ave; unit: deg) & 0 & 0.00073 & 0.00840 & 0.00731 & 0.00717\\
 & SNR (ave) & 25.910 & 25.910 & 25.940 & 42.559 & 42.458\\
SKA1-LOW & Position error (ave; unit: deg) & 0 & 0.00074 & 0.00566 & 0.00298 & 0.00294\\
 & SNR (ave) & 57.476 & 57.591 & 59.050 & 75.967 & 76.027\\
SKA AA2 & Position error (ave; unit: deg) & 0 & 0.00073 & 0.00932 & 0.00375 & 0.00262\\
 & SNR (ave) & 23.460 & 23.460 & 23.456 & 47.569 & 47.365\\
\hline
\end{tabular}
\end{table*}

From the perspective of source detection sensitivity, TOI (both 2-D and 3-D) achieves a SNR substantially higher than WSClean and SKA-SDP across all three telescope configurations, with a consistent improvement of at least 15 in SNR. From the perspective of source position accuracy, TOI is most directly comparable to WSClean (\textit{w}-snapshot), as both avoid full \textit{w}-stacking correction and operate under a snapshot imaging framework. Relative to WSClean (\textit{w}-snapshot), TOI achieves a significant improvement in source position accuracy; even with TOI (2-D), the position error is reduced by 12.98\%, 47.35\%, and 59.76\% compared to WSClean (\textit{w}-snapshot) for MeerKAT, SKA1-LOW, and SKA AA2, respectively. Furthermore, compared to WSClean (\textit{w}-snapshot), TOI (3-D) reduces the position error by 14.64\%, 48.06\%, and 71.89\% for the three telescopes, respectively. It shows that the advantage of TOI is particularly evident for larger arrays, such as SKA AA2, where the 3-D TOI approach effectively resolves the increased non-coplanarity associated with longer baselines. 

Nevertheless, the sensitivity improvement (higher SNR) achieved by TOI comes at the expense of position accuracy compared to full \textit{w}-stacking methods, such as WSClean ($w$-stacking) and the SKA-SDP imager. To mitigate this limitation, as described in Section \ref{FIPsec52}, FIP-TOI uses TOI (imager) and FITrig (transient finder) to identify candidate pulsar tiles, while the final pulsar position is measured from a full-resolution difference image reconstructed by the SKA-SDP imager. More precisely, this difference image is generated from the two time-adjacent snapshots exhibiting the strongest pulsar intensity variation over the observation period, as identified through the \texttt{tLISI} cube analysis. The SKA-SDP imager is run only once, on these two selected snapshot indices, to produce the high-precision full-resolution difference image used for the final position measurement. A detailed evaluation of this behaviour is presented in Section \ref{FIP553}.

\subsection{Performance of pipelines}
\label{FIP552}

\subsubsection{Estimation on simulated datasets}

\begin{figure}
    \centering
    \includegraphics[width=0.9\columnwidth]{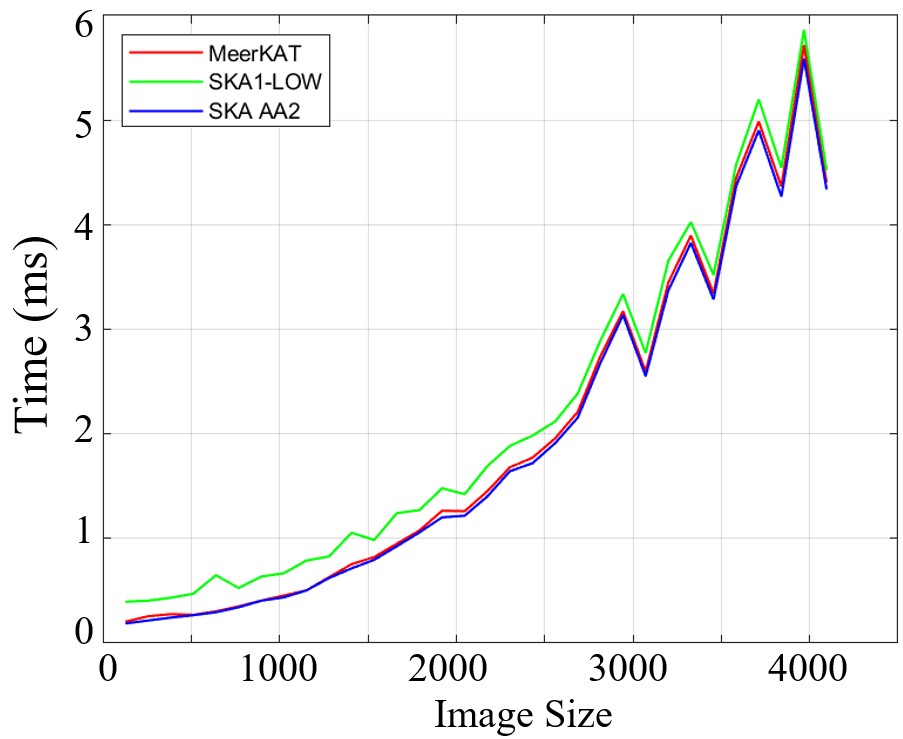}
\caption{Computation time of unfused kernels in TOI (2-D) on one snapshot, with image size represented by $N$ pixels.
\label{pipetime}}
\end{figure}

In our experiments, FIP-TOI is executed on a NVIDIA H100\footnote{\url{https://resources.nvidia.com/en-us-tensor-core}}. To assess computational efficiency for data generated from various telescope configurations, the TOI processes images of varying sizes, conducting 10 trials per size. By measuring the execution time of the CUDA imaging kernels and averaging across all 10 runs, we derive the results presented in Fig. \ref{pipetime}.

In our experiments, the single-frequency Measurement Sets for a single time slot contain 130816 baselines for SKA1-LOW, 2080 for SKA AA2, and 1711 for MeerKAT. The results indicate that variations in telescope configurations have little impact on the computation time among these simulated datasets, as the visibilities are handled in parallel.

\begin{figure}
    \centering
    \includegraphics[width=\columnwidth]{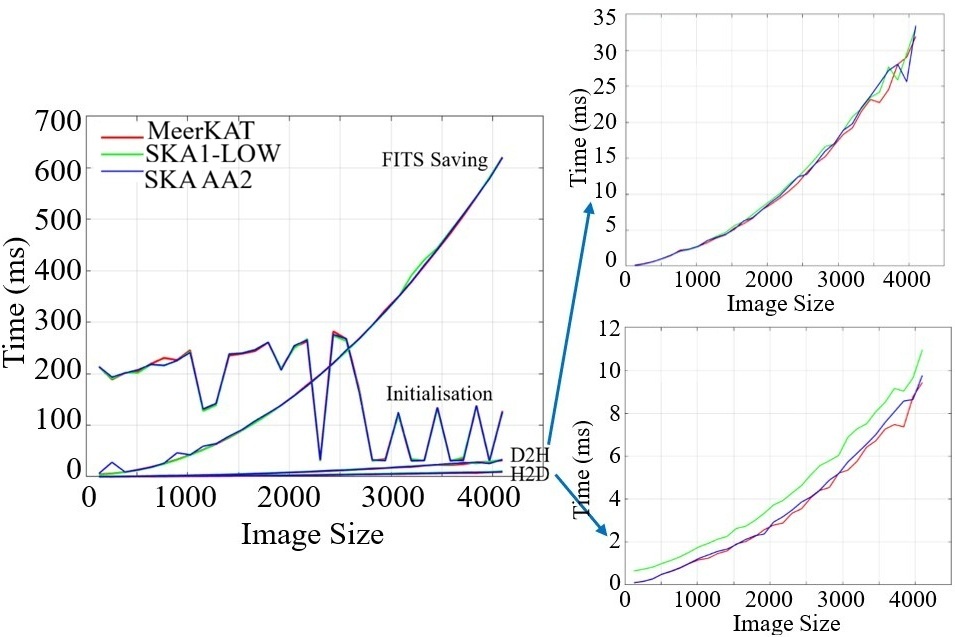}
    \caption{Execution time for initialisation, PCIe host-to-device (H2D), PCIe device-to-host (D2H), and FITS saving in the TOI, with image size represented by $N$ pixels. The red, green, and blue curves illustrate the performance on data simulated for MeerKAT, SKA1-LOW, and SKA AA2, respectively.
\label{timesepa}}
\end{figure}

When running the TOI as a stand-alone imaging tool, concurrent bidirectional transfer can be used. Once data transfer is finalised, the dirty snapshot is stored as a FITS file using the C/C++ interface. The execution time for each operation, averaged over 10 trials, is illustrated in Fig. \ref{timesepa}.

\begin{figure}
    \centering
    \includegraphics[width=0.95\columnwidth]{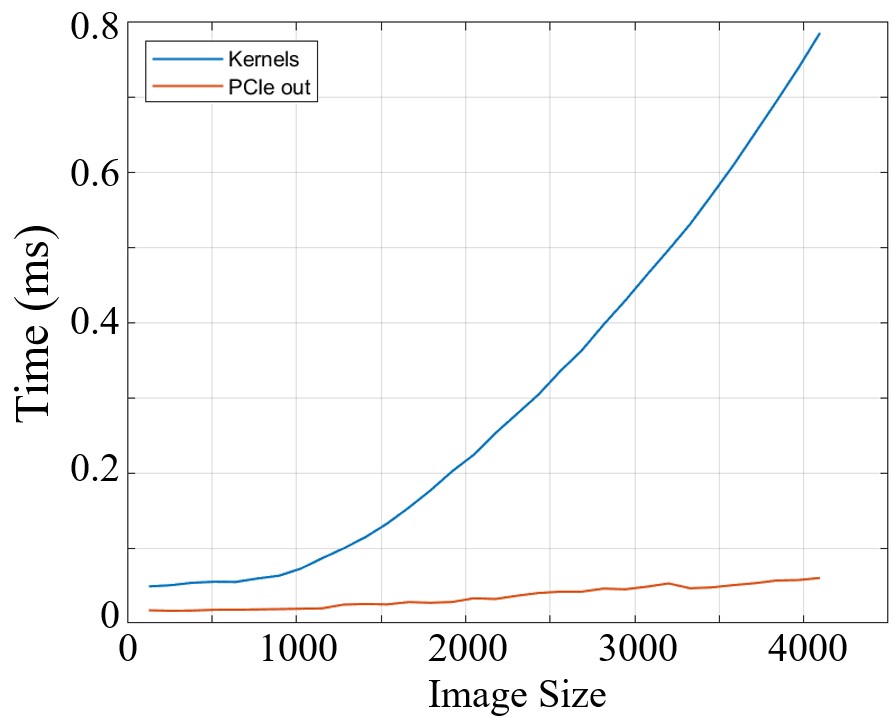}
\caption{Execution time of kernels and PCIe D2H within FITrig, with the image size represented by $N$ pixels.
\label{triggertimesmall}}
\end{figure}

When integrated into the FIP, TOI is directly linked to FITrig, which takes dirty snapshots as input. This eliminates the need to transfer them (D2H/H2D) back and forth through the PCIe bus, or save them as intermediate FITS files. The FITrig kernels initiate processing once three consecutive snapshots have been produced by the TOI. At the end of the process, only the tLISI matrix is transferred (D2H). For reference, Fig. \ref{triggertimesmall} illustrates the FITrig's kernel execution time and D2H transfer time for different image sizes. Since the FITrig works solely with pixel intensities, variations in telescope configurations do not impact its computation time.

\begin{figure}
\centering
    \begin{subfigure}
    {\columnwidth}\centering\includegraphics[width=0.95\columnwidth]{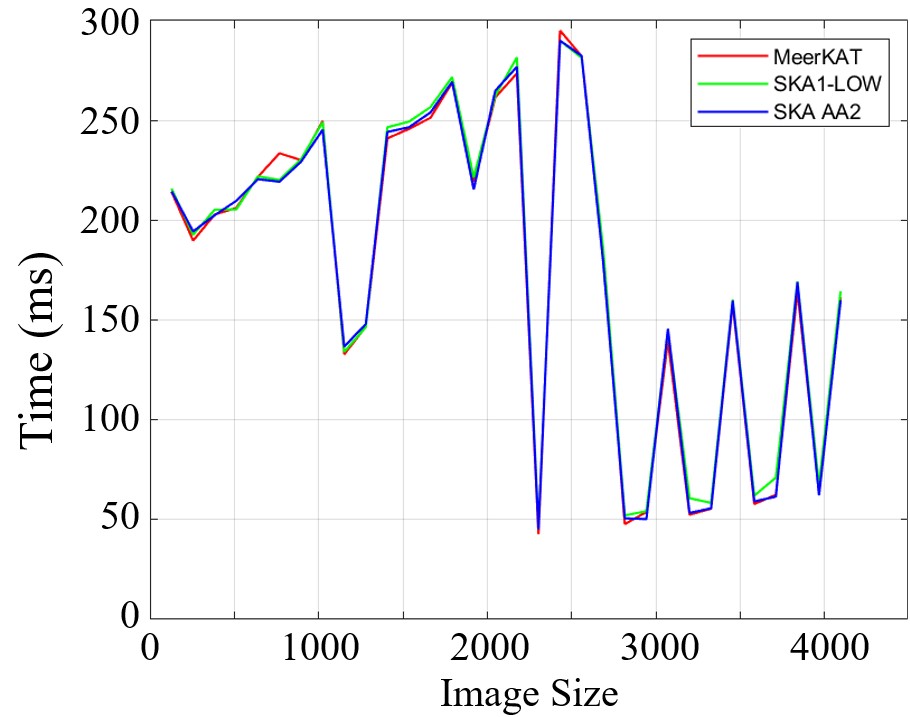}\subcaption{}\end{subfigure}
    \begin{subfigure}{\columnwidth}\centering\includegraphics[width=0.95\columnwidth]{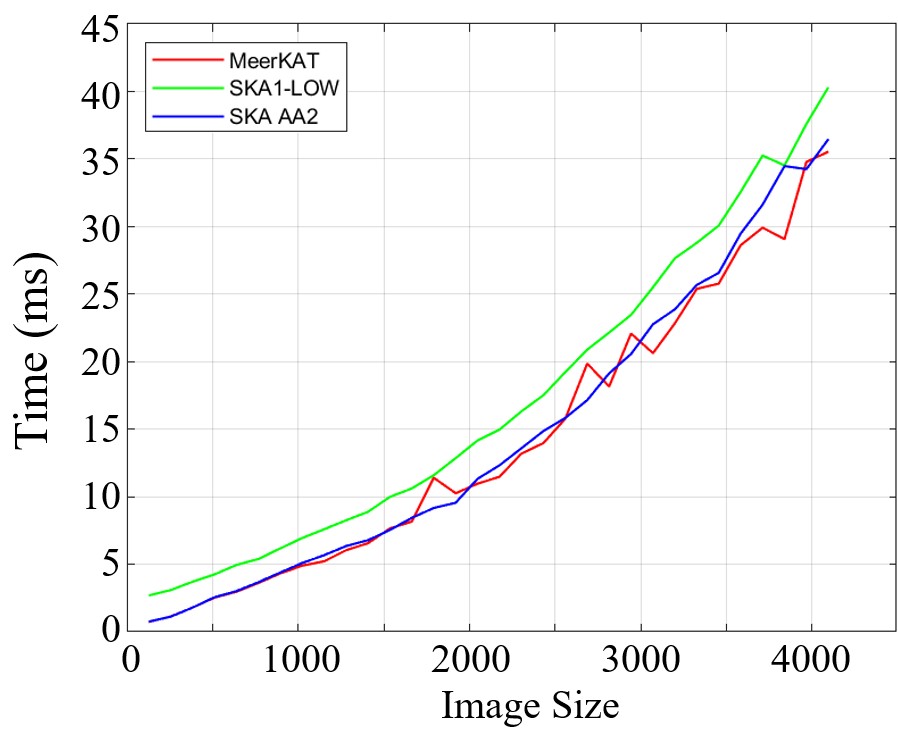}\subcaption{}\end{subfigure}
\caption{FIP-TOI computation time for data simulated with various telescope configurations, where (a) includes the initialisation time and (b) omits the \texttt{cufftPlan2d} time, with image size represented by $N$ pixels.
\label{svdoverall}}
\end{figure}

The average time required to produce a single tLISI matrix with FIP-TOI, based on 10 trials, is presented in Fig. \ref{svdoverall} (a). The most time-consuming component of this process is the \texttt{cufftPlan2d}, which is 2-D FFT configuration included in the initialisation. When the TOI produces consecutive snapshots of identical image size and baseline counts, whether as a stand-alone imaging tool or as a component in the FIP, GPU initialisation procedures such as \texttt{cudaMalloc} and \texttt{cufftPlan2d} are executed only once at the start. As the number of computed tLISI matrices increases, the effect of initialisation time on each individual computation becomes less significant. Therefore, if we exclude the time spent on \texttt{cufftPlan2d}, the remaining computation time for FIP-TOI is illustrated in Fig. \ref{svdoverall} (b). As shown, different telescope configurations have minimal influence on the FIP-TOI's computation time.

For comparison, we analyse the execution time of the WSClean-based FIP, where WSClean functions as the imaging component of the FIP and FITrig functions as the localisation component. Here in this approach, the dirty snapshots produced by WSClean need to be H2D transferred prior to executing the FITrig. WSClean supports multiple gridding methods. In this study, we analyse its performance using both the default $w$-stacking gridder and the $w$-snapshot gridder. In experiments utilising the $w$-stacking gridders, we activate the \texttt{-intervals-out} option to enhance WSClean's efficiency by enabling automatic separable computation for each snapshot. Each dataset, generated based on different telescope layouts, consists of ten snapshots. To determine the average processing time per snapshot, we divide the total runtime by 10.

For experiments involving the $w$-snapshot gridder, WSClean first applies phase rotation to the snapshot data using \texttt{chgcentre -minw}, then executes the imaging process while shifting back the phase to the original target by external re-projection module (not timed here). We conduct 10 trials and compute the average time.

\begin{figure}
    \centering
    \begin{subfigure}{0.8\columnwidth}\includegraphics[width=\linewidth]{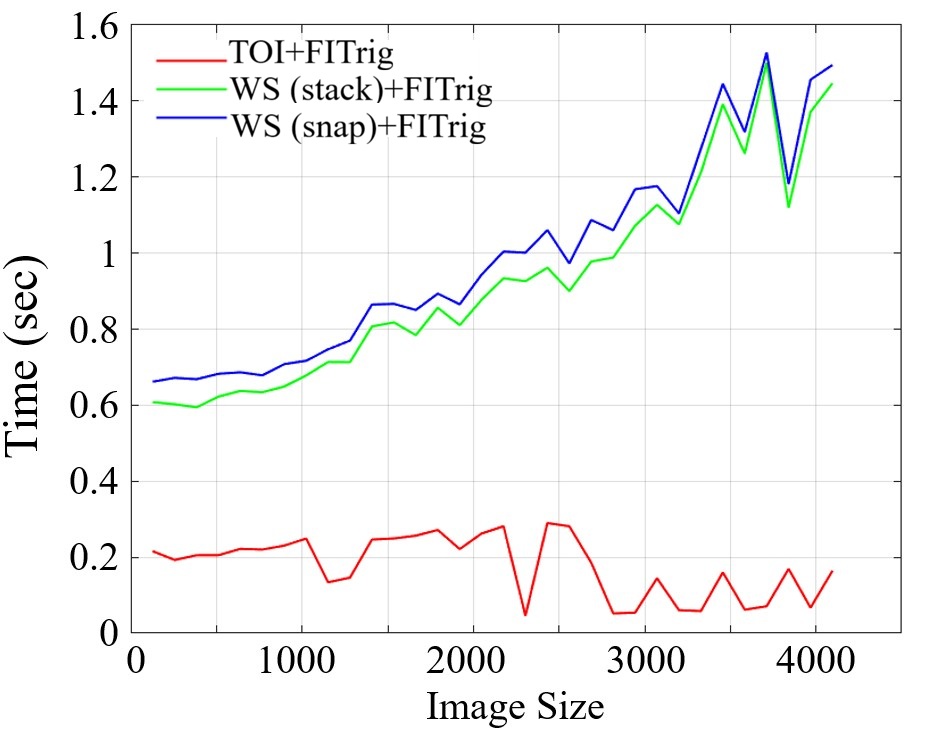}\subcaption{}\end{subfigure}
    \begin{subfigure}{0.8\columnwidth}\includegraphics[width=\linewidth]{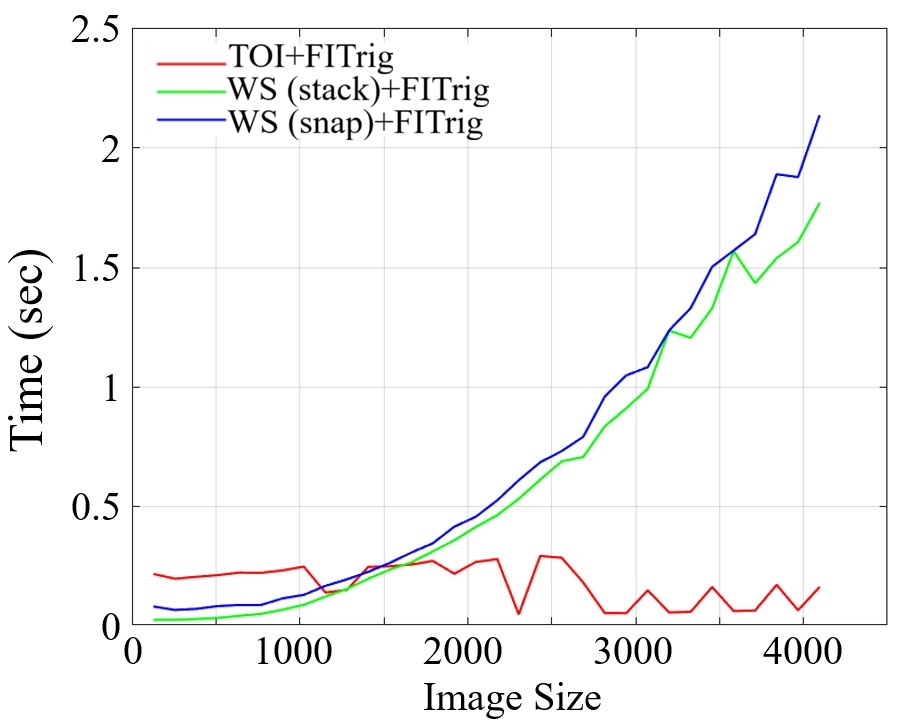}\subcaption{}\end{subfigure}
    \begin{subfigure}{0.8\columnwidth}\includegraphics[width=\linewidth]{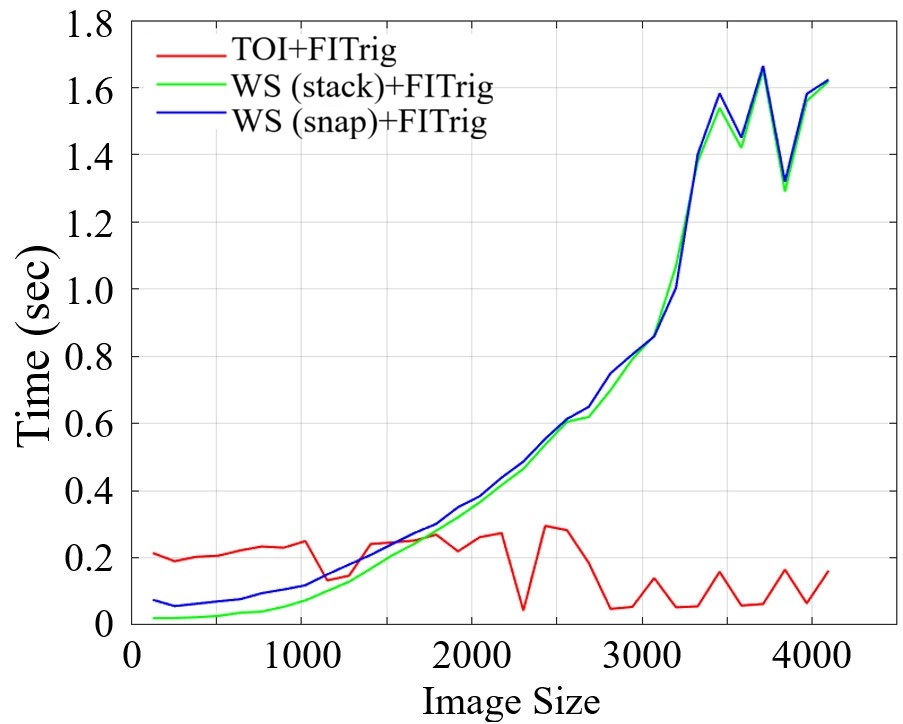}\subcaption{}\end{subfigure}
\caption{Computation time comparison for FIP-TOI and WSClean-based pipelines, with results displayed separately for datasets simulated with (a) SKA1-LOW, (b) SKA AA2, and (c) MeerKAT telescope layouts, with image size represented by $N$ pixels.
\label{svdoverallcomp}}
\end{figure}

Figure \ref{svdoverallcomp} presents a comparison of computation times for FIP-TOI and WSClean-based pipelines using $w$-stacking or $w$-snapshot gridders on datasets simulated with different telescope layouts.

As shown in Fig. \ref{svdoverallcomp}, WSClean's $w$-snapshot and per-snapshot implementation of $w$-stacking exhibit similar performance. However, its performance differs depending on the telescope configuration. This variation arises because different configurations lead to differences in baseline counts and visibility coverage, influencing the required gridding and iFFT operations.

For MeerKAT and SKA AA2, the WSClean-based pipeline processes data more quickly when dealing with smaller images (below approximately $1.8\mathrm{K} \times 1.8\mathrm{K}$ pixels). In contrast, for larger images exceeding this size, our FIP-TOI outperforms the WSClean-based pipeline. Quantitatively, our FIP-TOI operates considerably faster than the WSClean-based pipeline, delivering speed improvements of 8.80, 11.08, and 10.03 times for 4K $\times$ 4K-pixel images produced by datasets from SKA1-LOW, SKA AA2, and MeerKAT telescope configurations, respectively.

%
%
%
%
 
\subsubsection{Timing and resource usage on a large-scale real dataset}

In order to evaluate the performance of the highly parallelised FIP-TOI implemented in Section \ref{FIP54}, we conduct timing and computational resource usage characterisation on a real Measurement Set (1500 snapshots, 1024 frequency channels, 1770 baselines per snapshot) of PSR J0901-4046 collected by MeerKAT. The dataset is 369 GiB (396 GB) in size. Five different image sizes (2048, 4096, 6144, 8192, 10240) and three different FOV ( 1\textdegree, 2\textdegree and 3\textdegree) are tested, with each combination tested at least six times for a total of 90 trials.

The hardware for this characterisation is a single-socket Lenovo ThinkSystem SD665-N V3 Neptune DWC Server with one AMD EPYC 9454 48-Core CPU (2.75 GHz Base, 3.8 GHz Boost) and four NVIDIA H100 GPUs (80 GB) hosted at Digital Research Alliance Canada's ``Fir'' cluster.

Overall processing is found to be largely invariant across the configurations tested, with wall times of 170-190 seconds in all trials. This corresponds to an average processing rate of 1500/180 $\approx$ 8.33 snapshots per second, or 120 milliseconds per snapshot.

Resource utilisation is similarly insensitive to the configuration. RAM and CPU resources required are found to be low, and nearly completely invariant to the configuration tested: 714 MB RAM and approximately 2 CPU cores are used continuously.

Detailed timing measurements instead indicate that the overall runtime is dominated by FITS file writing (the \texttt{data\_write} stage in Section \ref{sec41}). The CFITSIO library universally used for FITS file access makes use only of C standard I/O routines for I/O, which default in Glibc to 4-8 KiB buffer sizes regardless of filesystem. No CFITSIO API exists to override it. At the maximum image size tested (10240), \texttt{data\_write} I/O still represents 90\% of the runtime. The output FITS cube is 615 MB and at a speed of 190 seconds this amounts to a write speed of no more than $\approx$ 3.4 MB/s. Despite this, the amount of data ingested is immense. A minimum of 109 GB of data must have been read, and at 180 seconds this represents 608 MB/s.

\begin{figure}
    \includegraphics[width=0.9\columnwidth]{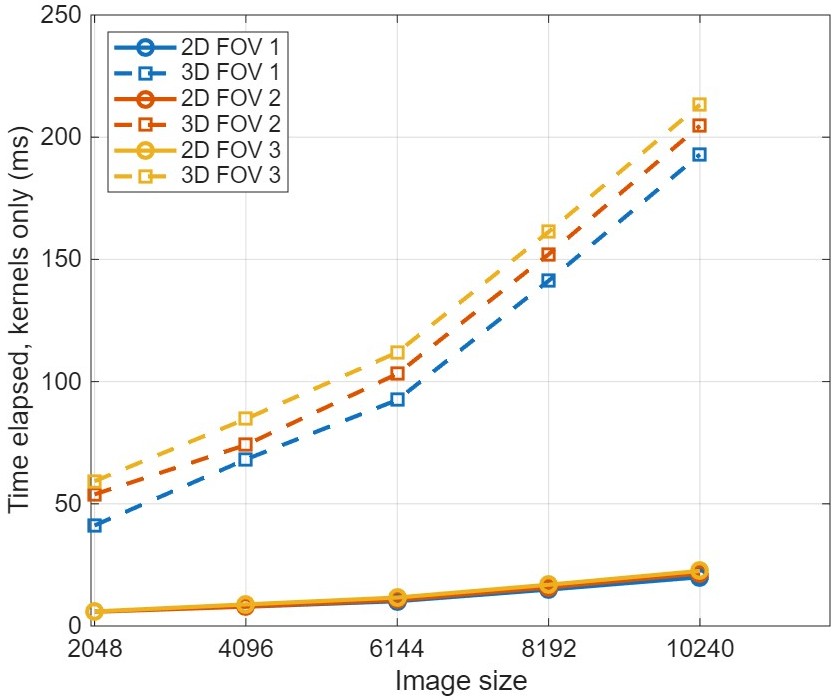}
\caption{Kernel execution time per snapshot for TOI (2-D) and TOI (3-D) on a single A100 GPU across different image sizes ($N$) and FOVs (degrees).\label{2d3d}}
\end{figure}

For comparison, SKA-SDP-FIP requires more than 256 GB of RAM to process the same 1500-snapshot Measurement Set with image size of 4096 $\times$ 4096 pixels and FOV of 1.7 degrees. The quickest run on a single NVIDIA A100 GPU (80GB) in our experiment is 1.225 hours. In contrast, with comparable configuration and running on the same GPU, FIP-TOI (2-D) has significantly smaller requirements in time (13.5 minutes), CPU (1-2 CPU cores) and memory (<1 GB). It achieves this by never retaining more than $O(1)$ state (eight snapshots) in memory at any time, efficiently scheduling data transfers to and from the GPU, and heavily optimising the processing on the GPU. Figure \ref{2d3d} illustrates the kernel execution time for TOI (2-D) and TOI (3-D) on a single A100 GPU. Even when replacing the 2-D implementation with the 3-D version for higher precision, the execution time will remain short and is still substantially lower than that of SKA-SDP-FIP.

\section{Discussions}
\label{FIP553}

In this section, we present and analyse the performance of FIP-TOI in scenarios with different pulsar behaviours, including FOV with multiple pulsars (Section \ref{gleam5}), an on-and-off pulsar (Section \ref{sectionpul}) and a pulsar exhibiting gradual intensity changes (Section \ref{pulsargrad}). The performance of FIP-TOI is compared with that of SKA-SDP-FIP.

\subsection{Performance on GLEAM}
\label{gleam5}

To verify the effectiveness of FIP-TOI on datasets with multiple transients, we simulate a Measurement Set (128 frequency channels, 130816 baselines per snapshot) based on the GLEAM catalogue \citet{gleam}. The Measurement Set includes three snapshots, which is the minimum number for FIP pulsar search \citet{FItrigger}. The first two snapshots employ the sky model with sources from ``sky2.osm'' in \citealt{dataMS}, while the third snapshot employs the sky model with sources from ``sky1.osm'' in \citealt{dataMS}. The main difference between them is that the five faintest sources present in \textit{sky1} are missing in \textit{sky2}; these sources are used to represent transients. While it is rare for 5 transients within the same FOV to vanish simultaneously during real-world observations, this experiment aims to showcase the pipelines' performance in handling situations with multiple transients in a single FOV.

\begin{figure}
    \centering
    \begin{subfigure}{0.45\columnwidth}\centering\includegraphics[width=\linewidth]{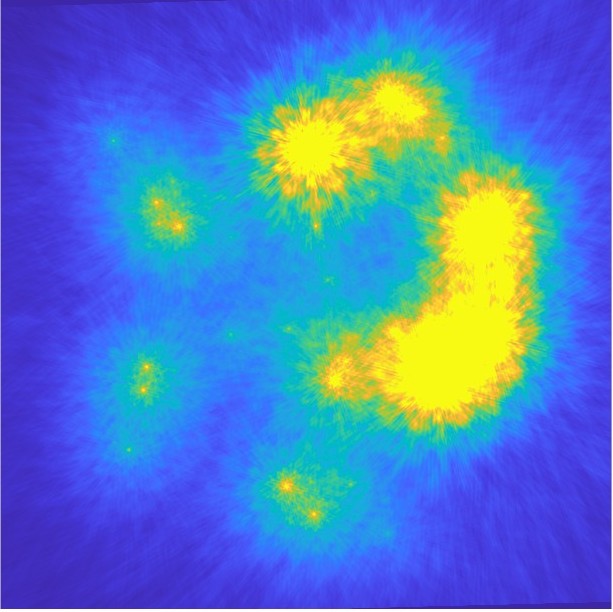}\subcaption{}\end{subfigure}
        \begin{subfigure}{0.535\columnwidth}\centering\includegraphics[width=\linewidth]{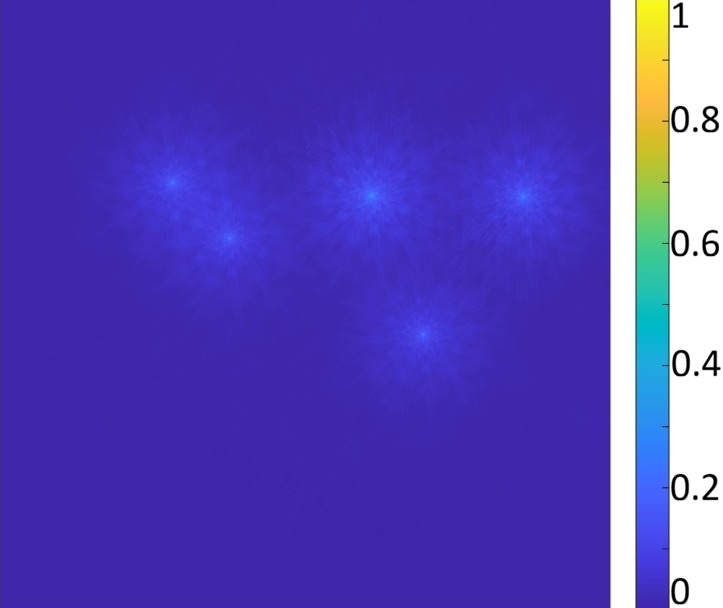}\subcaption{}\end{subfigure}
    \begin{subfigure}{0.45\columnwidth}\centering\includegraphics[width=\linewidth]{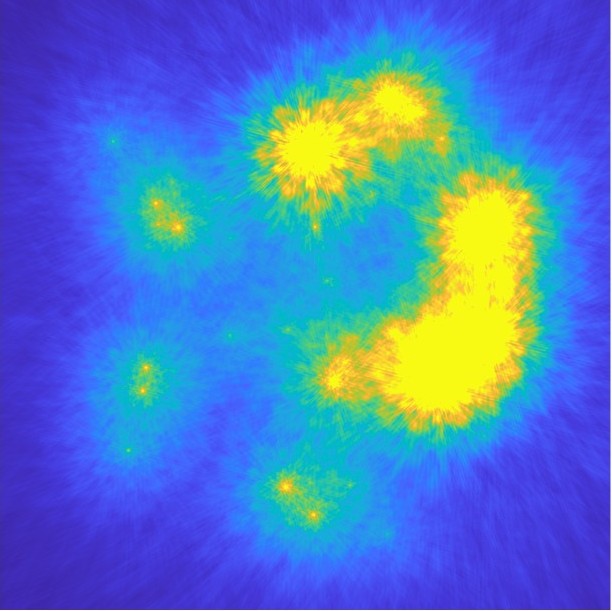}\subcaption{}\end{subfigure}
        \begin{subfigure}{0.535\columnwidth}\centering\includegraphics[width=\linewidth]{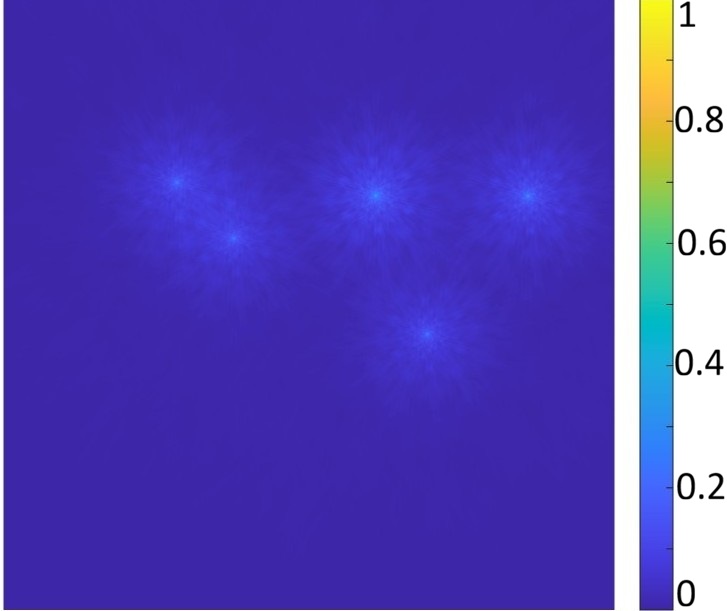}\subcaption{}\end{subfigure}
    \caption{GLEAM snapshots and difference images (with logarithmic representation): (a) the third snapshot generated by TOI (2-D), (b) the difference between the second and third snapshots generated by TOI (2-D); (c) the third snapshot generated by SKA-SDP imager; and (d) the difference between the second and third snapshots generated by SKA-SDP imager. Images are normalised for illustration purpose.\label{gleamimg}}
\end{figure}

Since the GLEAM Catalogue\footnote{\url{https://heasarc.gsfc.nasa.gov/W3Browse/radio-catalog/gleamegcat.html}} is an extragalactic catalogue observed by the MWA \citet{mwa}, an SKA1-LOW precursor, we adopt the SKA1-LOW layout in simulations, with a 1-second dump time \citet{ska1low}. The third dirty snapshots produced by SKA-SDP imager and TOI are illustrated in Fig. \ref{gleamimg}, with a 1.7-degree FOV. The five faint ``transients'' are visible in the difference images. The images are sized 4096 $\times$ 4096 pixels. The tile size in FITrig is selected to be $32 \times 32$ pixels \citet{FItrigger}.

The detection results produced by both FIPs are illustrated in Table \ref{gleam61}. The transient positions used in the simulation serve as ground truths for the detection. The position error is calculated as the Euclidean distance between the detected and ground-truth positions in the RA/Dec coordinate space.

\begin{table*}
\centering
\caption{Detection results of simulated transients in GLEAM, where the position of each transient candidate is shown as RA, Dec in the unit of degree. Ground truth fluxes are listed in the unit of Jy.\label{gleam61}}
\begin{tabular}{ccccc}
 \hline
\# & \multicolumn{3}{c}{Position} & Flux\\
 & Ground Truth & SKA-SDP-FIP & FIP-TOI & \\
 \hline
1 & 19.784298, -30.302931 & 19.783680, -30.302798 & 19.783849, -30.302725 & 0.0464\\
2 & 19.294044, -30.299221 & 19.293402, -30.299003 & 19.293638, -30.299259 & 0.0410\\
3 & 20.426094, -30.338543 & 20.425579, -30.338394 & 20.425479, -30.338545 & 0.0321\\
4 & 19.620411, -29.916634 & 19.619802, -29.916447 & 19.619958, -29.916651 & 0.0317\\
5 & 20.241962, -30.183567 & 20.241507, -30.183640 & 20.241751, -30.183683 & 0.0302\\
\hline
\multicolumn{2}{l}{Average Error} & 0.00058883 & 0.00044218 \\
\multicolumn{2}{l}{Maximum Error} & 0.00067800 & 0.00061500 \\
\hline
\end{tabular}
\end{table*}

The results demonstrate that both FIPs correctly detect the sources, while FIP-TOI provides slightly higher accuracy. The cell size in this experiment is 0.000415 degree, which is comparable to the detection errors. The flux of the brightest stationary source in the FOV is 0.523 Jy, and the five sources absent have fluxes between 0.030 Jy and 0.046 Jy, making them up to 17.4 times dimmer than the brightest stationary source. This experiment highlights the capability of FIP-TOI in detecting multiple faint transients with high precision.

\subsection{Performance on PSR J0901-4046}
\label{sectionpul}

To verify the performance of FIP-TOI on datasets with an on-and-off pulsar, we carry out experiment on the Measurement Set (1500 snapshots, 1024 frequency channels, 1770 baselines per snapshot) of PSR J0901-4046 collected by MeerKAT, with a dump time of 1.999 seconds.

\begin{figure}
    \centering
    \begin{subfigure}{0.45\columnwidth}\centering\includegraphics[width=\linewidth]{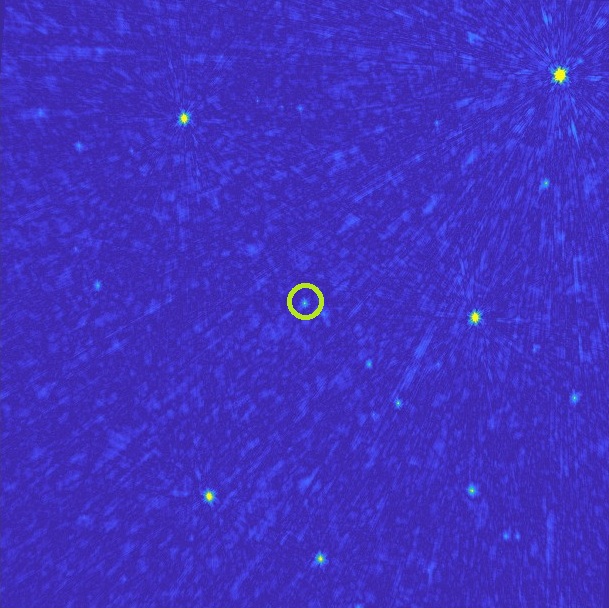}\subcaption{}\end{subfigure}
    \begin{subfigure}{0.535\columnwidth}\centering\includegraphics[width=\linewidth]{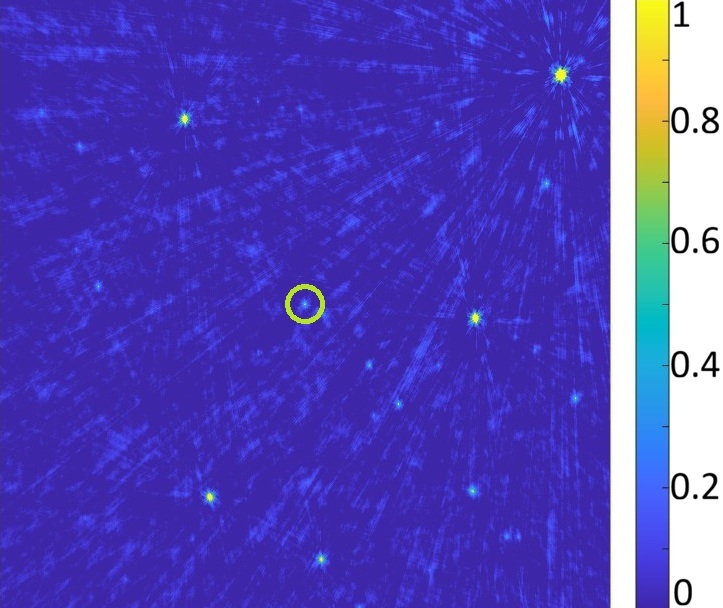}\subcaption{}\end{subfigure}
    \caption{Dirty snapshots showing PSR J0901-4046, visible to the human visual system, reconstructed with (a) TOI (2-D) and (b) SKA-SDP imager. The pulsar is marked with circles. Images are normalised for illustration purpose and shown using a logarithmic representation.
\label{psr09}}
\end{figure}

As an example, Fig. \ref{psr09} presents a dirty snapshot (1.2-degree FOV) in which the pulsar is perceptible to the human visual system for both pipelines. The intensity scales in both sub-figures are normalised for illustration purposes, with the colour bar representing the normalised pixel intensities. The images are 4096 $\times$ 4096 pixels in size, where the cell size is set to 0.000293 degrees.

In FIP-TOI, by applying FITrig to dirty snapshots obtained by TOI with a tile size of $32 \times 32$ pixels, a \texttt{tLISI} cube is obtained. The subsequent detection and position estimation procedure follows that described in Section \ref{FIPsec52}. In particular, candidate detections are identified from the Fourier transformed \texttt{tLISI} cube and converted into bounding boxes, within which the pulsar position is subsequently estimated from a high-accuracy full-resolution difference image.

Within each detected bounding box, the local median background is subtracted and source pixels are selected using a local $5\sigma$ threshold. The connected component containing the brightest pixel is then required to be consistent with a Gaussian-like source morphology before its intensity-weighted centroid is used as the source position. This criterion is appropriate for the higher-frequency MeerKAT images, where compact sources are expected to be better represented by the synthesised beam. In contrast, the SKA1-Low case in Section \ref{gleam5} uses a simpler centroiding procedure without the Gaussian-like morphology requirement, since lower-frequency images can have broader and more complex source responses, making such a constraint overly restrictive.

The detection results produced by both FIPs are illustrated in Table \ref{psr0901}. The results show that both pipelines correctly detect the desired pulsar, while FIP-TOI shows higher precision.

\begin{table}
\centering
\caption{Detection result of PSR J0901-4046, where the position is shown as RA, Dec in the unit of degree.\label{psr0901}}
\begin{tabular}{cccc}
 \hline
Coordinate & Ground Truth & SKA-SDP-FIP & FIP-TOI\\
 \hline
RA & 135.371871 & 135.371341 & 135.371673\\
Dec & -40.767496 & -40.767863 & -40.767740\\
\hline
\multicolumn{2}{l}{Position Error} & 0.00064466 & 0.00031423\\
\hline
\end{tabular}
\end{table}

\subsection{Performance on PSR J1703-4902}
\label{pulsargrad}

\begin{figure}
    \centering
    \includegraphics[width=\columnwidth]{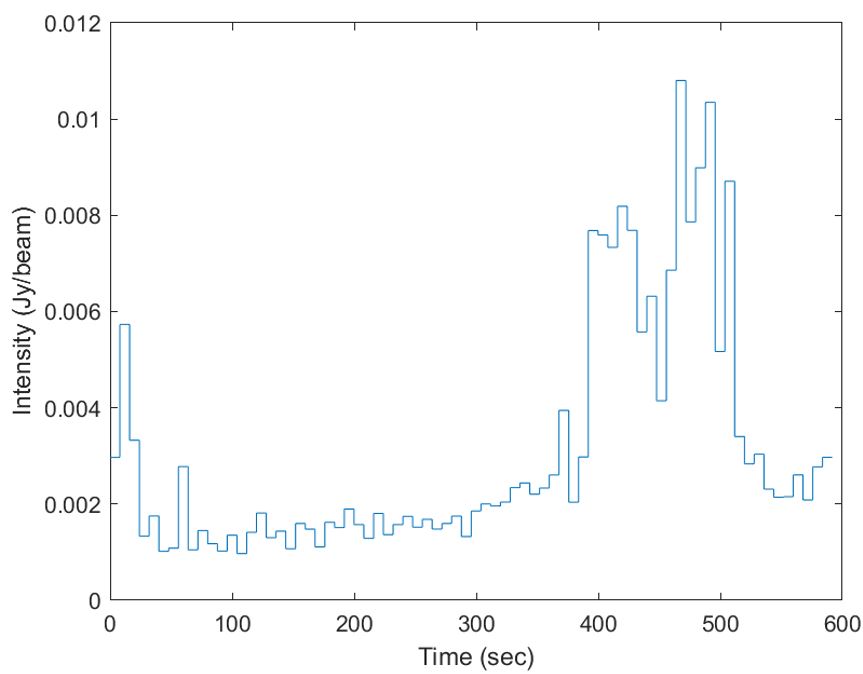}
\caption{Intensity fluctuation of PSR J1703-4902 in the measurement.
\label{profile}}
\end{figure}

In contrast to PSR J0901-4046, which shows an on-and-off pattern throughout the observation, PSR J1703-4902\footnote{\url{https://simbad.cds.unistra.fr/simbad/sim-id?bibyear1=1850&bibyear2=\%24currentYear&submit=Display&Ident=\%403509446&Name=PSR+J1703-4902&bibdisplay=refsum&bibyear1=1850&bibyear2=\%24currentYear}} \citet{psr1,psr2} exhibits a gradual variation in brightness over time, as depicted in Fig. \ref{profile}. This figure is derived by recording the peak pixel intensity of PSR J1703-4902 in each restored snapshot (within a 1.0-degree FOV), deconvolved using WSClean. The observation was carried out with MeerKAT, employing a dump time of 7.9966 seconds. The Measurement Set contains 74 snapshots, 1024 frequency channels, and 1891 baselines per snapshot.

\begin{figure}
    \centering
    \begin{subfigure}{0.455\columnwidth}\centering\includegraphics[width=\linewidth]{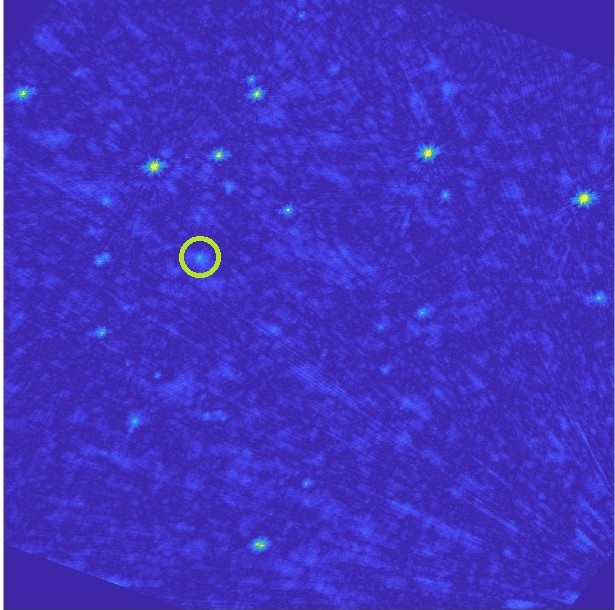}\subcaption{}\end{subfigure}
    \begin{subfigure}{0.535\columnwidth}\centering\includegraphics[width=\linewidth]{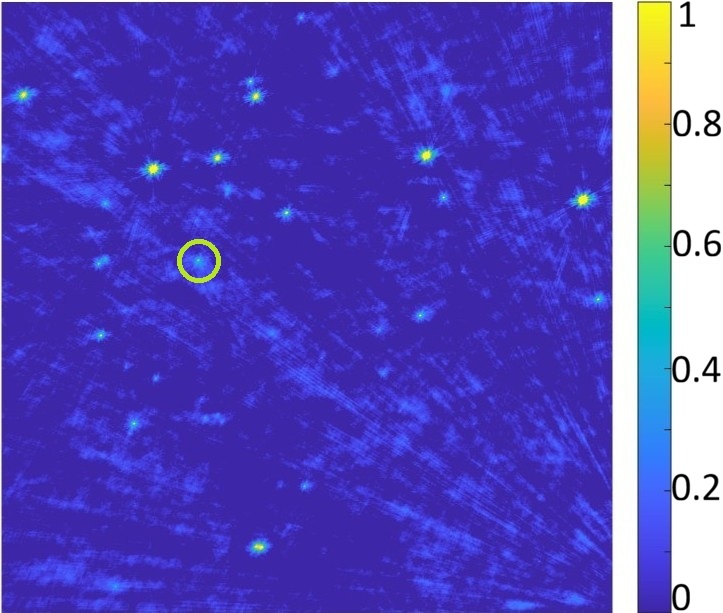}\subcaption{}\end{subfigure}
    \caption{Dirty snapshots showing PSR J1703-4902, visible to the human visual system, reconstructed with (a) TOI (2-D) and (b) SKA-SDP imager. The pulsar is marked with circles. Images are normalised for illustration purpose and shown using a logarithmic representation.
\label{psrnew}}
\end{figure}

As an example, Fig. \ref{psrnew} presents a dirty snapshot in which the pulsar is perceptible to the human visual system for both pipelines. The intensity scales in both sub-figures are normalised for illustration purposes, with the colour bar representing the normalised pixel intensities. The images are 4096 $\times$ 4096 pixels in size, where the cell size is set to 0.000245 degrees.

In FIP-TOI, we apply FITrig to dirty snapshots reconstructed by TOI, using $32 \times 32$-pixel tiles, resulting in a \texttt{tLISI} cube. In this experiment, the Measurement Set we have contains only one period of the pulsar, so the pulsar does not provide multiple repeated cycles over which a periodic detection statistic can be coherently reinforced. In this case, applying a Fourier transform and harmonic summing to the \texttt{tLISI} cube is, therefore, not the most appropriate detection strategy. Instead, for each spatial tile the temporal anomaly score is computed by summing $1-\mathrm{tLISI}$ along the time axis, so that tiles whose similarity drops strongly during the transient receive larger scores. This 2-D score map is then converted into a z-score significance map and tiles that exceed the detection threshold are selected as candidate transient locations. Because the non-periodic feature does not benefit from coherent reinforcement through Fourier analysis, its detection statistic is expected to have lower contrast for faint transients. We therefore use a slightly lower $4\sigma$ threshold for the non-periodic significance map, rather than the $5\sigma$ threshold used for periodic searches, to preserve sensitivity to weak transient events. Candidate detections are still required to form significant connected tile regions and are subsequently verified through full-resolution difference-image localisation.

The same source positioning procedure described in Section \ref{sectionpul} is applied to the detected bounding boxes for the MeerKAT data of PSR J1703-4902. This includes local background subtraction, $5\sigma$ source-pixel selection, and the Gaussian-like morphology criterion.

The detection results produced by both FIPs are illustrated in Table \ref{psr1703}. The results show that both pipelines correctly detect the desired pulsar, while FIP-TOI still shows higher precision.
\begin{table}
\centering
\caption{Detection result of PSR J1703-4902, where the position is shown as RA, Dec in the unit of degree.\label{psr1703}}
\begin{tabular}{cccc}
 \hline
Coordinate & Ground Truth & SKA-SDP-FIP & FIP-TOI\\
 \hline
RA & 255.977250 & 255.976445 & 255.977038\\
Dec & -48.866944 & -48.866962 & -48.866750\\
\hline
\multicolumn{2}{l}{Position Error} & 0.00080520 & 0.00028737\\
\hline
\end{tabular}
\end{table}

The result is especially encouraging because the pulsar signal in this dataset is not an ideal periodic case: its intensity changes gradually across the observation, it is measured with $\sim 8$-second visibility dumps, which is a relatively long integration time, and it appears non-periodic to the detection pipeline because the data contain only one period. Even under these conditions, TOI maintains high sensitivity and localisation precision.

\section{Conclusion}
\label{FIP57}

Fast Imaging Pipeline is an umbrella term for image-based transient detection pipelines developed for real-time or quasi-real-time transient search with next-generation radio telescopes such as the SKA. Time-domain approaches can provide rapid transient detection, but often face challenges in accurately localising transient sources. In contrast, image-based FIP approaches operate directly on spatially resolved radio images, enabling precise transient localisation and thereby offering a significant advantage for the follow-up characterisation and observation of celestial transients.

The development of FIP within the SKA Science Data Processor (SKA-SDP-FIP) remains at an early design stage, leaving substantial scope to improve its imaging accuracy, computational efficiency, and scalability. In this work, we therefore propose a high-performance solution for high-precision and higher efficiency automated transient searches in radio images.

Specifically, we propose TOI, an innovative wide-field imaging technique designed for fast and accurate transient detection. Combined with FITrig, a high-sensitivity and high-efficiency transient finder developed in \citet{FItrigger}, TOI forms FIP-TOI, a highly parallelised pipeline designed for high-precision real-time pulsar localisation. The key contributions of this work are as follows:

\begin{itemize}
    \item \textbf{Transient-oriented imaging formulation:} TOI re-formulates dirty snapshot imaging around the instantaneous geometric structure of the baseline distribution, starting from the vector-based form of visibility equation. Using an SVD-derived coordinate system, the dominant visibility plane is imaged directly, while residual non-coplanarity can be retained through a near-coplanar Chebyshev correction, which is especially useful for large telescope arrays with long baselines. This formulation provides a snapshot-based wide-field imager that maintains high transient-search sensitivity and SNR while improving positional accuracy relative to $w$-snapshot imaging. At the same time, it avoids the computational cost of applying full $w$-correction to every snapshot, as required by $w$-stacking.
    \item \textbf{GPU-native re-projection and data handling:} The re-projection from the transformed image plane to the standard image plane is formulated as an algebraically reduced pixel-to-pixel mapping suitable for CUDA execution. This eliminates the need for CPU-oriented WCS routines in the main imaging path and enables independent pixels to be transformed in parallel. Combined with direct Measurement Set input, weighted polarisation handling, multi-frequency synthesis, flag handling, and in-device hand-off from TOI to the dirty-image-based FITrig, this design makes the implementation portable for wider astronomers by using standard data format as input, while reducing unnecessary host-device data transfers and avoiding the computational cost of deconvolution.
    \item \textbf{Integrated image-based transient detection and localisation:} FIP-TOI combines TOI (imager) with FITrig (transient finder) to identify candidates from temporal variations in dirty snapshot sequences, rather than applying a conventional source finder independently to each difference image. Candidate regions are firstly identified using the tLISI cube and subsequently refined through full-resolution difference-image localisation. This integration removes several catalogue construction and candidate-merging steps required by source-finding-based detection designs, reduces the number of user-configurable parameters, and provides a natural framework for streaming transient searches.
    \item \textbf{Highly parallelised radio astronomy workflow:} Systematically applying ring buffering, asynchronous execution, and fused operations to radio transient imaging pipeline, producing a GPU-native FIP-TOI design that is unusually well matched to the throughput and memory constraints of astronomical snapshot pipelines. It comes closer than any other FIP to perfectly mating the large computational requirements of radio astronomical image processing with the difficult-to-tap, immense power of GPUs, thereby enabling efficient processing of large data volumes and supporting real-time transient detection under SKA-scale observational conditions.
\end{itemize}

FIP-TOI achieves arcsecond-level localisation precision while providing efficient, low-resource processing. On a 396 GB MeerKAT Measurement Set, FIP-TOI processes 1500 snapshots on a single NVIDIA H100 in approximately 170-190 seconds across image size ranging from 2048 $\times$ 2048 to 10240 $\times$ 10240 pixels and FOVs ranging from 1 to 3 degrees, corresponding to an average processing time of 120 ms per snapshot. Resource usage is similarly insensitive to the configuration, with only 714 MB of RAM and approximately two CPU cores required continuously. This real-time processing capability and reduced memory footprint facilitate timely follow-up observations and enable processing close to the point of data acquisition, reducing the need to transfer large volumes of data between processing sites.

The pipeline is evaluated using both simulated SKA-layout datasets and MeerKAT observations of PSR J0901-4046 and PSR J1703-4902. These experiments cover several challenging scenarios, including multiple faint transients within a single FOV, an on-and-off long-period pulsar, and a pulsar exhibiting gradual temporal intensity variation that appears non-periodic to the detector because of truncated observation duration. Across these cases, FIP-TOI successfully detects the target transients and achieves lower positional errors than the SKA-SDP-FIP comparison pipeline, demonstrating its robustness beyond an idealised case of a strictly periodic pulsar.

It is important to note that the existing SKA-SDP-FIP remains a prototype at the current stage of SKA development. We hope that FIP-TOI can bring a breakthrough to the SKA mission. This potential impact is particularly timely given the SKAO's timeline towards science operations\footnote{\url{https://www.skao.int/en/science-users/timeline-science}}. In addition, although FIP-TOI is evaluated only on pulsar data in this work, it is promising for extension to other transient sources, such as FRBs, which we will investigate in future work.

\section*{Acknowledgements}

The authors acknowledge support from STFC Grant (ST/W001969/1). The authors acknowledge the use of the University of Oxford Advanced Research Computing (ARC; \citealt{arc}) facility in carrying out this work. This research was enabled in part by compute resources provided by Mila (\url{mila.quebec}). This research was enabled in part by support provided by WestGrid and the Digital Research Alliance of Canada (\url{alliancecan.ca}). The authors made use of GNU Parallel \citet{tange_2023_7958356} in the course of their work. Special thanks are extended to Ian Heywood for supporting the data of PSR J0901-4046 and to Rob Fender and Joe Bright for supporting the data of PSR J1703-4902. The authors are particularly grateful to Ali Taqi for helpful discussions on projection. Gratitude is extended to Fred Dulwich and Ben Mort for valuable guidance on OSKAR. The authors also express sincere thanks to Feng Wang and Yangfan Xie for their insightful discussions on \textit{ws}-snapshot. The authors are especially grateful to Mark Calabretta, Francois-Xavier Pineau, and Matthieu Baumann for helpful discussions on re-projection and WCSLIB. The authors' heartfelt thanks go to André Offringa for the enlightening discussions on WSClean. We wish to express our sincere thanks to Michael Giles for helpful knowledge and practice about CUDA programming learnt from the Course on CUDA Programming on NVIDIA GPUs (\url{https://people.maths.ox.ac.uk/gilesm/cuda/index.html}).

\section*{Data Availability}
Data supporting the findings of this study are available from the corresponding author upon request.

\bibliographystyle{mnras}
\bibliography{Bibliography.bib}

@article{aegean1, 
    title="{Source finding in the era of the SKA (Precursors): Aegean 2.0}", 
    volume={35}, 
    journal={PASA}, 
    author={Hancock, Paul J. and Trott, Cathryn M. and Hurley-Walker, Natasha}, 
    year={2018}, 
    pages={e011},
    doi={10.1017/pasa.2018.3}
}

@article{askap_vast, 
    title={The ASKAP Variables and Slow Transients (VAST) Pilot Survey}, 
    volume={38}, 
    DOI={10.1017/pasa.2021.44}, 
    journal={PASA}, 
    author={Murphy, Tara and Kaplan, David L. and Stewart, Adam J. and others}, 
    year={2021}, 
    pages={e054}
}

@article{CASAnew,
	Author = {{The CASA Team} and Ben Bean and Sanjay Bhatnagar and others},
	Journal = {PASP},
	Number = {114501},
	Pages = {},
	Title = "{CASA, the common astronomy software applications for radio astronomy}",
	Volume = {134},
	Year = {2022},
    doi={10.1088/1538-3873/ac9642}
}

@article{FItrigger,
  author={Xiaotong Li and Karel Adámek and Wesley Armour},
  journal={A\&C}, 
  title="{FITrig: A high-performance detection technique for efficient ultra-long-period pulsars}", 
  year={2026},
  volume={55},
  number={},
  pages={101058},
  doi={10.1016/j.ascom.2025.101058},
}

@ARTICLE{gleam,
author = {N. Hurley-Walker and J. Callingham and P. Hancock and others},
title = "{GaLactic and Extragalactic All-sky Murchison Widefield Array (GLEAM) survey - I. A low-frequency extragalactic catalogue}",
journal = {MNRAS},
year = {2017},
volume = {464},
number = {1},
pages = {1146-1167},
doi={10.1093/mnras/stw2337}
}

@article{mwa, 
title={The Murchison Widefield Array: The Square Kilometre Array Precursor at Low Radio Frequencies}, 
volume={30}, 
journal={PASA}, 
author={Tingay, S. J. and Goeke, R. and Bowman, J. D. and others}, 
year={2013}, 
pages={e007},
doi={10.1017/pasa.2012.007}
}

@article{nifty,
	author = {Selig, M. and Bell, M. R. and Junklewitz, H. and others},
	title = "{NIFTY --- Numerical Information Field Theory - A versatile PYTHON library for signal inference}",
	journal = {A\&A},
	year = 2013,
	volume = 554,
	pages = "A26",
	month = "",
    doi = {10.1051/0004-6361/201321236}
}

@article{psr1,
    author = {Driessen, L N and Stappers, B W and Tremou, E and others},
    title = "{21 new long-term variables in the GX 339−4 field: two years of MeerKAT monitoring}",
    journal = {MNRAS},
    volume = {512},
    number = {4},
    pages = {5037-5066},
    year = {2022},
    month = {03},
    doi={10.1093/mnras/stac756}
}

@article{psr2,
    author = {Tremou, E and Corbel, S and Fender, R P and others},
    title = "{Radio and X-ray detections of GX 339-4 in quiescence using MeerKAT and Swift}",
    journal = {MNRAS Letters},
    volume = {493},
    number = {1},
    pages = {L132-L137},
    year = {2020},
    month = {02},
    doi={10.1093/mnrasl/slaa019}
}

@article{realfast,
doi = {10.3847/1538-4365/aab77b},
year = {2018},
publisher = {The American Astronomical Society},
volume = {236},
number = {1},
pages = {8},
author = {Law, C. J. and Bower, G. C. and Burke-Spolaor, S. and others},
title = {realfast: Real-time, Commensal Fast Transient Surveys with the Very Large Array},
journal = {ApJS},
}

@article{realpul1,
	Author = {Pelisoli, I. and Marsh, T.R. and Buckley, D.A.H. and others},
	Journal = {Nat Astron},
	Number = {},
	Pages = {931-942},
	Title = "{A 5.3-min-period pulsing white dwarf in a binary detected from radio to X-rays}",
	Volume = {7},
	Year = {2023},
    doi={10.1038/s41550-023-01995-x}
}

@article{realpul2,
	Author = {Caleb, M. and Heywood, I. and Rajwade, K. and others},
	Journal = {Nat Astron},
	Number = {},
	Pages = {828-836},
	Title = "{Discovery of a radio-emitting neutron star with an ultra-long spin period of 76 s}",
	Volume = {6},
	Year = {2022},
    doi={10.1038/s41550-022-01688-x}
}

@ARTICLE{sfind,
       author = {{Hopkins}, A.~M. and {Miller}, C.~J. and {Connolly}, A.~J. and others},
        title = {A New Source Detection Algorithm Using the False-Discovery Rate},
      journal = {AJ},
         year = 2002,
        month = feb,
       volume = {123},
       number = {2},
        pages = {1086-1094},
        doi = {10.1086/338316}
}

@article{sofia2,
    author = {Westmeier, T and Kitaeff, S and Pallot, D and others},
    title = "{sofia 2 - an automated, parallel H i source finding pipeline for the WALLABY survey}",
    journal = {MNRAS},
    volume = {506},
    number = {3},
    pages = {3962-3976},
    year = {2021},
    month = {07},
    doi={10.1093/mnras/stab1881}
}

@article{sofia22,
    author = {Serra, Paolo and Westmeier, Tobias and Giese, Nadine and others},
    title = "{SoFiA: a flexible source finder for 3D spectral line data}",
    journal = {MNRAS},
    volume = {448},
    number = {2},
    pages = {1922-1929},
    year = {2015},
    month = {02},
    doi = {10.1093/mnras/stv079}
}

@article{sourcefinder1,
    author = {Hancock, P. J. and Murphy, T. and Gaensler, B. M. and others},
    title = "{Compact continuum source finding for next generation radio surveys}",
    journal = {MNRAS},
    volume = {422},
    number = {2},
    pages = {1812-1824},
    year = {2012},
    month = {04},
    doi={10.1111/j.1365-2966.2012.20768.x}
}

@article{transient2,
year = {2018},
month = {oct},
publisher = {Uspekhi Fizicheskikh Nauk, Russian Academy of Sciences and IOP Publishing},
volume = {61},
number = {10},
pages = {965},
author = {S B Popov and K A Postnov and M S Pshirkov},
title = {Fast radio bursts},
journal = {Phys.-Usp.},
doi={10.3367/UFNe.2018.03.038313}
}

@article{w3,
	Author = {Cornwell, T. and Golap, K. and Bhatnagar, S.},
	Journal = {IEEE J. Sel. Topics Signal Process.},
	Number = {},
	Pages = {647-657},
	Title = "{The noncoplanar baselines effect in radio interferometry: The W-Projection Algorithm}",
	Volume = {2},
	Year = {2008},
    doi={10.1109/JSTSP.2008.2005290}
}

@article{w6,
   Author = {Offringa, A. and McKinley, B. and Hurley-Walker, N. and others},
   Journal = {MNRAS},
   Number = {1},
   Pages = {606-619},
   Title = "{WSCLEAN: An implementation of a fast, generic wide-field imager for radio astronomy}",
   Volume = {444},
   Year = {2014},
   doi={10.1093/mnras/stu1368}
}

@article{wcs,
	author = {{Calabretta, M. R.} and {Greisen, E. W.}},
	title = "{Representations of celestial coordinates in FITS}",
	journal = {A\&A},
	year = 2002,
	volume = 395,
	number = 3,
	pages = "1077-1122",
    doi={10.1051/0004-6361:20021327}
}

@article{wcs1,
	author = {{Greisen, E. W.} and {Calabretta, M. R.}},
	title = "{Representations of world coordinates in FITS}",
	journal = {A\&A},
	year = 2002,
	volume = 395,
	number = 3,
	pages = "1061-1075",
    doi={10.1051/0004-6361:20021326}
}

@article{wgridder,
    author = {Ye, Haoyang and Gull, Stephen F and Tan, Sze M and Nikolic, Bojan},
    title = "{High accuracy wide-field imaging method in radio interferometry}",
    journal = {MNRAS},
    volume = {510},
    number = {3},
    pages = {4110-4125},
    year = {2021},
    month = {12},
    doi={10.1093/mnras/stab3548}
}

@article{wssnap,
    author = {Xie, Yang-Fan and Wang, Feng and Deng, Hui and others},
    title = "{WS-Snapshot: An effective algorithm for wide-field and large-scale imaging}",
    journal = {MNRAS},
    volume = {515},
    number = {2},
    pages = {1973-1981},
    year = {2022},
    month = {07},
    doi={10.1093/mnras/stac1843}
}

@book{Pulsar,
  title = "{Handbook of Pulsar Astronomy}",
  author = {D. Lorimer and M. Kramer},
  year = {2004},
  month = {December},
  publisher = {Cambridge University Press},
  series = {Cambridge Observing Handbooks for Research Astronomers}
}

@book{RAbook,
    author    = "Burke, B. and Graham-Smith, F. and Wilkinson, P.",
    title     = "{An Introduction to Radio Astronomy}",
    year      = "2019",
    publisher = "Cambridge University Press",
    address   = "Cambridge",
    Edition = "4"
}

@book{supnov,
  title = {Supernovae},
  edition = {2nd},
  author = {Paul Murdin and Lesley Murdin},
  year = {2011},
  month = {April},
  publisher = {Cambridge University Press}
}

@Inbook{vcz,
author="Thompson, A. Richard
and Moran, James M.
and Swenson, George W.",
title="Van Cittert--Zernike Theorem, Spatial Coherence, and Scattering",
bookTitle="Interferometry and Synthesis in Radio Astronomy",
year="2017",
publisher="Springer International Publishing",
address="Cham",
pages="767-786",
}

@inproceedings{CASA,
  author={J. McMullin and B. Waters and D. Schiebel and W. Young and K. Golap},
  booktitle="{ADASS XVI}", 
  title="{CASA architecture and applications}", 
  year={2007},
  volume={376},
  number={},
  pages={127-130},
  publisher = {ASP Conference Series},
  editor = {R. Shaw and F. Hill and D. Bell},
  series= {},
}

@inproceedings{CUDA,
    author = {Kirk, David},
    title = "{NVIDIA CUDA software and GPU parallel computing architecture}",
    booktitle = {Proceedings of the 6th International Symposium on Memory Management},
    year = {2007},
    publisher = {},
    editor = {},
    pages = {103-104},
    volume = {7},
    doi={10.1145/1296907.1296909}
}

@inproceedings{fipska,
  author = "Stolyarov, Vladislav  and  Li, Xiaotong  and  Heywood, Ian  and  Adamek, Karel",
  title = "{A fast imaging pipeline for transient detection in interferometric data}",
  doi = "10.22323/1.514.0009",
  booktitle = "PoS(HEASA2025)",
  year = 2025,
  volume = "514",
  pages = "009"
}

@INPROCEEDINGS{pulskasim,
  author={X. Li and V. Stolyarov},
  booktitle={ADASS XXXV}, 
  title="{PulSKASim: A pulsar simulator for SKA-scale interferometric observations}", 
  year={2025},
  volume={},
  number={},
  pages={},
  doi={10.48550/arXiv.2603.04593}}

@InProceedings{RA7,
    author = {Thompson, A.},
    title = {Fundamentals of radio interferometry},
    booktitle = "{Synthesis Imaging in Radio Astronomy II}",
    year = 1999,
    publisher = {Astronomical Society of the Pacific Conference Series},
    editor = {Taylor, G. and Carilli, C. and Perley, R.},
    pages = {11-36},
    Volume = 180,
    series= {},
}

@inproceedings{skaconf,
author = {Gerhard Swart and Philip Diamond and Lewis Ball and others},
title = {{Construction update for the Square Kilometre Array Observatory}},
volume = {13094},
booktitle = {Ground-based and Airborne Telescopes X},
editor = {Heather K. Marshall and Jason Spyromilio and Tomonori Usuda},
organization = {International Society for Optics and Photonics},
publisher = {SPIE},
pages = {130940L},
year = {2024},
series= {},
doi={10.1117/12.3021248}
}

@inproceedings{thunderkat,
author = {P. Woudt and R. Fender and S. Corbel and others},
title = {{ThunderKAT: The MeerKAT large survey project for image-plane radio transients}},
volume = {277},
booktitle = {MeerKAT Science: On the Pathway to the SKA (MeerKAT2016)},
publisher = {PoS},
pages = {},
year = {2018},
doi = {10.22323/1.277.0013},
}

@InProceedings{w4,
    author = {Cornwell, T. and Voronkov, M. and Humphreys, B.},
    title = "{Wide field imaging for the Square Kilometre Array}",
    booktitle = "{Image Reconstruction from Incomplete Data VII}",
    year = 2012,
    publisher = {SPIE},
    editor = {Bones, P. and Fiddy, M. and Millane, R.},
    pages = {},
    Volume = {8500},
    series={},
    doi={10.1117/12.929336}
}

@misc{arc,
  title = "{University of Oxford Advanced Research Computing}",
  DOI = {10.5281/zenodo.22558},
  author = "Richards, A",
  year = 2015
}

@misc{dataMS,
  title = "{Sky Models}",
  DOI = {10.5281/zenodo.14744218},
  author = "Xiaotong Li",
  publisher = {Zenodo},
  year = 2024
}

@misc{OSKAR1,
  author = {Dulwich, Fred and Mort, Benjamin and Stolyarov, Vladislav and others},
  title = "{OxfordSKA/OSKAR}",
  DOI = {10.5281/zenodo.5722575},
  publisher={Zenodo},
  year = {2022}
}

@misc{ska1low,
  author = {P. Dewdney and J. Wagg and R. Braun and W. Turner},
  title = "{SKA1-LOW Configuration - Constraints \& Performance Analysis}",
  howpublished ={\url{https://www.skao.int/sites/default/files/documents/d17-SKA-TEL-SKO-0000557_01_-DesignConstraints-1.pdf}},
  year = 2016
}

@misc{skaaa2,
  author       = {Sridhar, S. and Breen, Shari and Arumugam, V. and others},
  title        = "{A year in the life of the SKA telescopes: overview and main outcomes (SKAO-TEL-0002665, Revision 01)}",
  year         = {2025},
  publisher    = {Zenodo},
  doi          = {10.5281/zenodo.16950982},
}

@misc{skaarray,
  author       = {Sridhar, S. and {Science Operations,} and Breen, Shari and others},
  title        = "{SKAO staged delivery, array assemblies and layouts (SKAO-TEL-0002299 Revision 04)}",
  year         = {2025},
  publisher    = {Zenodo},
  doi          = {10.5281/zenodo.16951020},
}

@phdthesis{xiaotongthesis,
  author       = {Xiaotong Li}, 
  title        = {Image quality assessment-based fast imaging pipeline for transient detection with GPU acceleration},
  school       = {University of Oxford},
  year         = 2024,
  doi          = {10.5287/ora-g7nnok9o9}
}

@misc{tange_2023_7958356,
      author       = {Tange, Ole},
      title        = {GNU Parallel 20230522 ('Charles')},
      month        = May,
      year         = 2023,
      note         = {{GNU Parallel is a general parallelizer to run
                       multiple serial command line programs in parallel
                       without changing them.}},
      publisher    = {Zenodo},
      doi          = {10.5281/zenodo.7958356},
      url          = {https://doi.org/10.5281/zenodo.7958356}
}

\appendix
\section{Mathematical Foundation of WCS Re-Projection Process}
\label{appendix1}

This appendix summarises the SIN projection in standard WCS implementation \footnote{\url{https://github.com/Punzo/wcslib/blob/master/C/prj.c}}. In their papers \citet{wcs,wcs1}, these mathematical equations are not explicitly written, so we summarise them here for the readers' convenience and to serve as the basis of optimisations in Section \ref{sec33}.

The traditional formulation proceeds through intermediate angular coordinates, native spherical coordinates, a tilt correction, and finally pixel coordinates. The process begins by mapping pixel coordinates $(p_1,p_2)$ on the tilted plane to intermediate coordinates $(x,y)$:
\begin{equation}
\left\{ 
\begin{matrix}
x = -\frac{\pi}{180}\beta\left( p_1 - \left(\frac{N}{2}+1\right)\right)\\
y = \frac{\pi}{180}\beta\left(p_2-\left(\frac{N}{2}+1\right)\right) ~~\\
\end{matrix}
\right . ,
\end{equation}
where the image consists of $N \times N$ pixels and $\beta$ is the cell size in degrees. The coordinates $(x,y)$ are angular coordinates in radians.

Next, $(x,y)$ are converted into native spherical coordinates $(\phi,\theta)$. The value of $\phi$ is given by
\begin{equation}
    \phi = \left \{ \begin{matrix}
        0, ~~~~~~~~~~~~~~x = 0~\mathrm{and}~y \leq 0 \\
        180, ~~~~~~~~~~~x = 0~\mathrm{and}~y > 0 \\
        90, ~~~~~~~~~~~~y = 0~\mathrm{and}~x > 0 \\
        -90, ~~~~~~~~~~y = 0~\mathrm{and}~x < 0 \\
        \frac{180}{\pi}\mathrm{atan2}\left(x,-y\right), ~\mathrm{otherwise} \\
    \end{matrix} \right .
\end{equation}
The value of $\theta$ depends on the condition of $x^2+y^2$. For $x^2+y^2<0.5$, $\theta$ is calculated as
\begin{equation}
    \theta = \left \{ \begin{matrix}
        0,~~~~~~~~~~\sqrt{x^2+y^2}\geq 1~\mathrm{and}~\sqrt{x^2+y^2}-1<\epsilon\\
        90,~~~~~~~~~~~~~~~~~~~~~~~~~~~~~~~~~~~~~~~~~~~~\sqrt{x^2+y^2} = 0\\
        180,~~\sqrt{x^2+y^2} \leq -1~\mathrm{and}~\sqrt{x^2+y^2}+1>-\epsilon\\
        \frac{180}{\pi}\mathrm{acos}\left(\sqrt{x^2+y^2}\right),~~~~~~~~~~~~~~~~~~~~~\mathrm{otherwise}\\
    \end{matrix} \right . ;
\end{equation}
for $0.5 \leq x^2+y^2 \leq 1$, $\theta$ is determined using
\begin{equation}
    \theta = \left\{ \begin{matrix}
        -90,~\sqrt{1-(x^2+y^2)} \leq -1~\mathrm{and}~\sqrt{1-(x^2+y^2)}+1>-\epsilon\\
        0,~~~~~~~~~~~~~~~~~~~~~~~~~~~~~~~~~~~~~~~~~~~~~~~~~~~~~\sqrt{1-(x^2+y^2)}=0\\
        90,~~~~~~~~~\sqrt{1-(x^2+y^2)} \geq 1~\mathrm{and}~\sqrt{1-(x^2+y^2)}-1<\epsilon\\
        \frac{180}{\pi}\mathrm{asin}\left(\sqrt{1-(x^2+y^2)}\right),~~~~~~~~~~~~~~~~~~~~~~~~~~~~~~~~\mathrm{otherwise}\\
    \end{matrix}\right. ,
\end{equation}
where $\epsilon = 10^{-10}$ \citet{wcs}.

The third step involves transforming $(\phi,\theta)$ back to $(x,y)$, while accounting for the tilt angles $\xi$ and $\eta$ of the $t_1,t_2$-plane relative to the $l,m$-plane, where
\begin{equation}
        \xi = \frac{V_{31}}{V_{33}}, ~
        \eta = \frac{V_{32}}{V_{33}}
\end{equation}
The conversion is defined by
\begin{equation}
    \left\{ \begin{matrix}
        x = \frac{180}{\pi}\left(\cos{\theta}\sin{\phi}+\xi\left(1-\sin{\theta}\right)\right)~~\\
        y = -\frac{180}{\pi}\left(\cos{\theta}\cos{\phi}-\eta\left(1-\sin{\theta}\right)\right)
    \end{matrix}\right. .
\end{equation}

Finally, $(x,y)$ is mapped to pixel coordinates $(p'_1,p'_2)$ on the image plane, which is parallel to the $l,m$-plane. This step is performed using
\begin{equation}
    \left \{ \begin{matrix}
        p'_1 = -\frac{x}{\beta} + \frac{N}{2} + 1\\
        p'_2 = \frac{y}{\beta} + \frac{N}{2} + 1~~\\
    \end{matrix}\right . .
\end{equation}

At this stage, the dirty image has been fully re-projected onto the celestial sphere and mapped onto the standard image plane.

\section{Derivations in Support of Near Coplanar Scenario}

\subsection{Derivation of Equation (\ref{eqnt3})}
\label{appB1}
From Equation (\ref{equ9}), we have
  \begin{align}
  l &= V_{11}t_1 + V_{21}t_2 + V_{31}t_3, \label{B1} \\
  m &= V_{12}t_1 + V_{22}t_2 + V_{32}t_3, \label{B2} \\
  n - 1 &= V_{13}t_1 + V_{23}t_2 + V_{33}t_3 . \label{B3}
  \end{align}
  Substituting $n$ in Equation (\ref{B3}) with the direction-cosine relation $n = \sqrt{1-l^2-m^2}$, we obtain
  \begin{equation}
  \sqrt{1-l^2-m^2}
  =
  1 + V_{13}t_1 + V_{23}t_2 + V_{33}t_3 .
  \end{equation}
  Squaring both sides gives
  \begin{equation}
  1-l^2-m^2
  =
  \left(1 + V_{13}t_1 + V_{23}t_2 + V_{33}t_3\right)^2 .
  \end{equation}
  Substituting the expressions for $l$ and $m$ with Equations (\ref{B1}) and (\ref{B2}) yields
  \begin{align}
  1
  &-
  \left(V_{11}t_1 + V_{21}t_2 + V_{31}t_3\right)^2
  -
  \left(V_{12}t_1 + V_{22}t_2 + V_{32}t_3\right)^2 \notag \\
  &=
  \left(1 + V_{13}t_1 + V_{23}t_2 + V_{33}t_3\right)^2 .
  \end{align}
  After expanding and collecting terms, this becomes
  \begin{align}
  0
  &=\left(V_{11}^2+V_{12}^2+V_{13}^2\right)t_1^2
  + \left(V_{21}^2+V_{22}^2+V_{23}^2\right)t_2^2 \notag \\
  &\quad
  + \left(V_{31}^2+V_{32}^2+V_{33}^2\right)t_3^2
  + 2\left(V_{11}V_{21}+V_{12}V_{22}+V_{13}V_{23}\right)t_1t_2 \notag \\
  &\quad
  + 2\left(V_{11}V_{31}+V_{12}V_{32}+V_{13}V_{33}\right)t_1t_3 \notag \\
  &\quad
  + 2\left(V_{21}V_{31}+V_{22}V_{32}+V_{23}V_{33}\right)t_2t_3 \notag \\
  &\quad
  + 2V_{13}t_1 + 2V_{23}t_2 + 2V_{33}t_3 .\label{B7}
  \end{align}
  Since the rows and columns of $\mathbf{V}$ are basis vectors and $\mathbf{V}$ is orthogonal, Equation (\ref{B7}) reduces to
  \begin{equation}
  t_1^2 + t_2^2 + t_3^2
  + 2V_{13}t_1 + 2V_{23}t_2 + 2V_{33}t_3
  = 0 .
  \end{equation}
  Completing the square gives
  \begin{equation}
  (t_1+V_{13})^2 + (t_2+V_{23})^2 + (t_3+V_{33})^2
  =
  V_{13}^2+V_{23}^2+V_{33}^2 =1.
  \end{equation}
  Solving for $t_3$ gives
  \begin{equation}
  t_3
  =
  -V_{33}
  \pm
  \sqrt{
  1
  -
  (t_1+V_{13})^2
  -
  (t_2+V_{23})^2
  }.
  \end{equation}

The branch is selected so that $t_3$ remains close to zero near the imaging plane, yielding
\begin{equation}
    t_3(t_1,t_2)=-V_{33} + \mathrm{sgn}(V_{33}) \sqrt{1-(t_1+V_{13})^2-(t_2+V_{23})^2}.
\end{equation}

\subsection{Chebyshev expansion of the residual phase}
\label{appB2}

According to Equation (\ref{eqn12}), the residual phase can be separated as
\begin{equation}
    e^{i2\pi r_{3,q}t_3}=e^{i2\pi r_{3,p}t_3}e^{i\alpha s_q}, ~\text{where} ~\alpha = 2 \pi \Delta r_3 t_3 .
\end{equation}

Since $s_q \in [-1,1]$, write $s_q=\cos \delta$. The Jacobi-Anger identity gives
\begin{equation}
    e^{i\alpha \cos \delta} = J_0(\alpha)+2\sum^\infty_{k=1}i^kJ_k(\alpha)\cos(k\delta).
\end{equation}
Using the Chebyshev identity
\begin{equation}
    T_k(s_q)=\cos(k\arccos s_q),
\end{equation}
we obtain
\begin{equation}
    e^{i\alpha s_q}=J_0(\alpha)+2\sum^\infty_{k=1}i^kJ_k(\alpha)T_k(s_q).
\end{equation}
The truncated expansion used by TOI is therefore
\begin{equation}
    e^{i\alpha s_q}\simeq \sum_{k=0}^{K-1}c_k(\alpha)T_k(s_q),
\end{equation}
where 
\begin{equation}
    \begin{cases}
        c_0(\alpha)=J_0(\alpha)\\
    c_k(\alpha)=2 i^k J_k(\alpha), k>0
    \end{cases} .
\end{equation}
Substitution into the near-coplanar dirty image expression yields the Chebyshev moment-image form used in Section \ref{sec34}.

\section{Additional figure(s) and table(s)}
\begin{figure*}
    \centering
    \includegraphics[angle=270, width=0.8\textwidth]{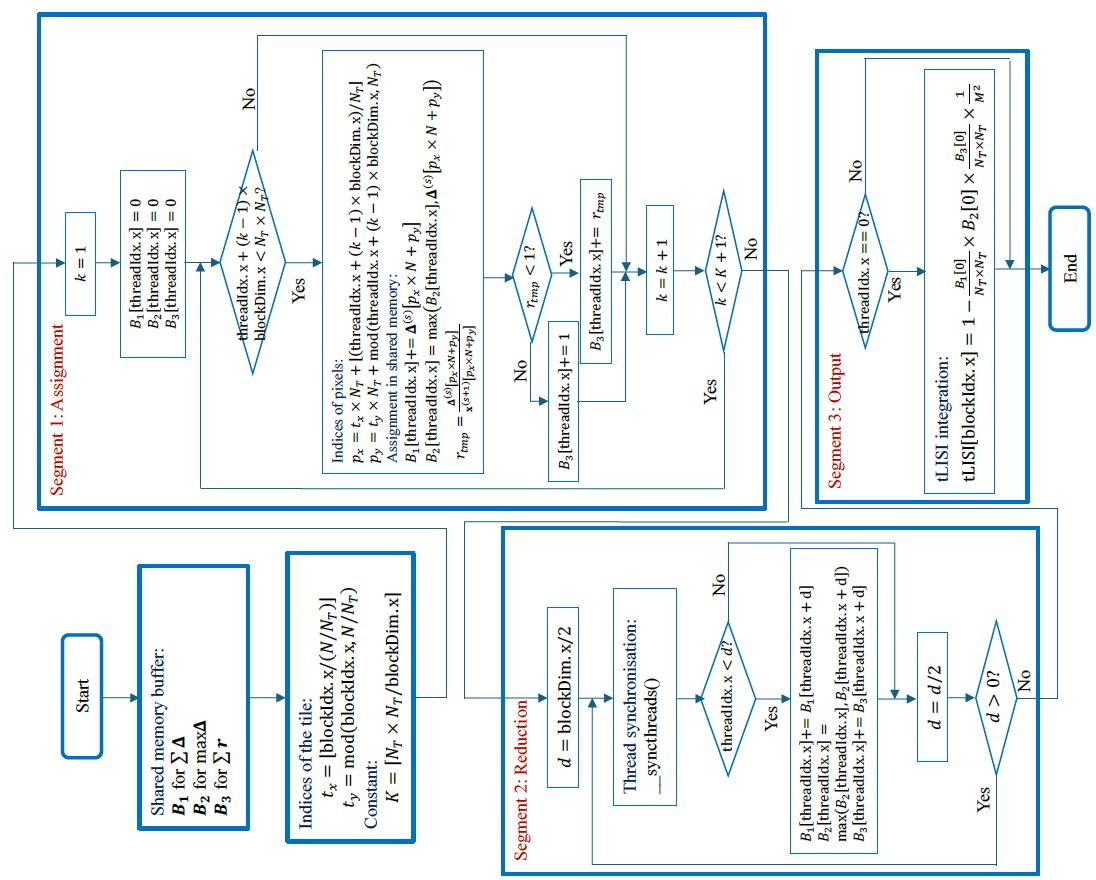}
\caption{Functional block diagram of tLISI kernel in FITrig.
\label{functlisi}}
\end{figure*}

\begin{itemize}
    \item \textbf{Figure \ref{functlisi}:} Functional block diagram of tLISI kernel in FITrig.
    \item \textbf{Table \ref{simupara}:} Simulation parameters used in Section \ref{FIP551}.
\end{itemize}

\begin{table*}
\centering
\caption{Simulation parameters used in Section \ref{FIP551}.\label{simupara}}
\begin{tabular}{cccc}
 \hline
Parameter & MeerKAT & SKA1-LOW & SKA AA2 (MID)\\
 \hline
Time per snapshot & 2 sec & 1 sec & 0.125 sec\\
Time-average duration & 2 sec & 1 sec & 0.125 sec\\
Number of frequency channels & 128 & 128 & 128\\
Start frequency & 1.2 GHz & 60 MHz & 1.2 GHz\\
Frequency increment & 1 MHz & 100 KHz & 1 MHz\\
Channel width & 1 MHz & 100 KHz & 1 MHz\\
Phase centre (RA, Dec) & 0 deg, 0 deg & 0 deg, 0 deg & 0 deg, 0 deg\\
Start time (UTC) & 2000-01-01 15:00:00 & 2000-01-01 12:00:00 & 2000-01-01 14:00:00\\
\hline
\end{tabular}
\end{table*}

\bsp	
\label{lastpage}
\end{document}